\documentclass[aps,showpacs]{revtex4}
\usepackage{amsmath}
\usepackage{graphicx}
\begin{document}

\title{Quantization of the Myers-Pospelov model: the photon sector
interacting with standard fermions as a perturbation of QED.}
\author{C. M. Reyes, L. F. Urrutia and J. D. Vergara}
\affiliation{Instituto de Ciencias Nucleares, Universidad Nacional Aut{\'o}noma de M{\'e}%
xico, \\
A. Postal 70-543, 04510 M{\'e}xico D.F., M{\'e}xico }

\begin{abstract}
We study the quantization of the electromagnetic sector of the Myers-Pospelov
model coupled to standard fermions. Our main objective, based
upon experimental and observational evidence, is to construct an effective
theory which is a  genuine perturbation of QED, such that setting zero
the Lorentz invariance violation parameters will reproduce it. To this end
we provide a physically motivated prescription, based on the effective
character of the model, regarding the way in which the model should be
constructed and how the QED limit should be approached. This amounts to the
introduction of an additional coarse-graining physical energy scale $M$,
under which we can trust the effective field theory formulation. The
prescription is successfully tested in the calculation of the Lorentz invariance 
violating contributions
arising from the electron self-energy. Such radiative corrections turn out to be properly scaled by
very small factors for any reasonable values of the parameters and no
fine-tuning problems are found. Microcausality violations are highly
suppressed and occur only in a space-like region extremely close to the
light-cone. The stability of the model is guaranteed by restricting to
concordant frames satisfying $1-|\mathbf{v}_{max}|> 6.5\times10^{-11}$.
\end{abstract}

\maketitle

\section{Introduction}

The Myers-Pospelov (MP) model \cite{MP} is an effective field theory
that incorporates scalars, fermions and photons in a particle (active)
Lorentz invariance violating (LIV) theory. It includes dimension five
operators, together with the presence of a fixed time-like direction $
n^{\mu }$ selecting a preferred frame. Such direction is assumed to arise
from a spontaneous Lorentz symmetry breaking in an underlying theory and
endows the model with covariance under observer (passive) Lorentz
transformations. The modified free Lagrangian density is
\begin{eqnarray}
&&\mathcal{L}_{MP}=-\phi ^{\ast }\left( \partial ^{2}+m^{2}\right) \phi +i%
\frac{\rho }{\tilde{M}}\phi ^{\ast }\left( n^{\mu }\partial _{\mu }\right)
^{3}\phi -\frac{1}{4}F_{\mu \nu }F^{\mu \nu }+\frac{\xi }{2 \tilde{M}}\left(
n^{\mu }F_{\mu \nu }\right) \left( n^{\alpha }\partial _{\alpha }\right)
\left( n_{\rho }\epsilon^{\rho\nu\kappa\lambda}F_{\kappa\lambda}\right)
\notag \\
&&+\;\;\bar{\Psi}i\gamma ^{\mu }\left( \partial _{\mu }-m\right) \Psi +\frac{%
1}{\tilde{M}}\bar{\Psi}\left( n_{\alpha }\gamma ^{\alpha }\right) \left(
\eta _{1}+\eta _{2}\gamma _{5}\right) \left( n^{\mu }\partial _{\mu }\right)
^{2}\Psi ,  \label{MPLAG}
\end{eqnarray}%
to which we add the electromagnetic interaction via the standard minimal
coupling. Such an effective theory is interpreted here as a model to
describe the imprints at Standard Model energies of active LIV, codified by
the dimensionless parameters $\rho, \xi, \eta_1$ and $\eta_2$, which is
produced by drastic modifications of the space-time structure at a
fundamental scale ${\bar M}$, as suggested by some phenomenological models inspired upon
developing quantum
gravity theories \cite{QGMODELS1,QGMODELS2,AMUFERMIONS} and string theories \cite{QGMODELS3}.
Nevertheless, up to now there is no
systematic derivation of a semiclassical approximation starting from
a fundamental  quantum gravity theory, for example, that could determine the exact
nature of the possible corrections arising from such space
granularity. This situation has prompted the construction and
analysis of effective field theories models which capture the basic
ingredients that we expect to survive at Standard Model energies.

The additional Lorentz violating terms in (\ref{MPLAG}) are unique according
to the following criteria: (i) quadratic in the same field, (ii) one more
derivative than the corresponding kinetic term, (iii) being gauge invariant,
(iv) being Lorentz invariant, except for the appearance of $n_\mu$, (v) not
reducible to lower dimension by the equations of motion and (vi) not
reducible to a total derivative \cite{MP}. The model has recently been
generalized to the non-abelian case including interactions arising from the
fields associated to the Standard Model \cite{BP}. As such, it could be
considered as a dimension-five-operator generalization of the Standard Model
Extension \cite{KOSTELECKY1}. In this work we will concentrate upon the
simpler version of Ref.\cite{MP}, particularly upon the proposed modified
electrodynamics in its quantized version. The corresponding classical model
has been thoroughly studied in relation to synchrotron radiation in Refs. %
\cite{MU}. Also, the self energy corrections of the model have been recently
analyzed in \cite{GV}. Radiative corrections to LIV theories have been studied in Refs.
\cite{RADCORR} and fine tuning problems have been discussed in Refs. \cite{GV,COLLINS}.

The point of view adopted in this work is to consider the quantum effective MP model
(\ref{MPLAG}) plus the electromagnetic interaction as a perturbation of the
Lorentz invariant theory, in the precise sense that after making zero the
LIV parameters encoding the corrections we must recover standard QED.
Moreover, since all experimental and observational evidence point to
negligible LIV \cite{MATTINGLY}, the radiative corrections arising from LIV
should be accordingly very small. As we will see in the sequel, this basic
idea provides a guideline in the way one gives a meaning to the model,
particularly in regard to its quantization and to the limiting procedure
necessary to recover QED.

Generally speaking, the dimension five operators  make the
theory of the higher order time-derivative (HOTD) type. This fact shows up
in the Lagrangian (\ref{MPLAG}) by the presence of third order time
derivatives for the scalars, second order time derivatives for the fermions
and third order time derivatives for the photons. It is well known that HOTD theories pose
many difficulties for their
implementation \cite{JAEN}, \cite{SIMON}, the most representative ones being
the increase in the number  of degrees of freedom with respect to the standard ones,
together with the appearance of Hamiltonians which are not positive definite
being unbounded from below. In this way, if one requires to treat the
additional HOTD terms as a perturbation, a careful strategy is required.
Fortunately, a systematic approach to carry out this task already exists in
the literature \cite{ELIEZER}, \cite{CHENGETAL}.

In view of the above considerations a general strategy to define the quantum
field theory extension of the MP model would be the following: (i) as usual,
the starting point is the classical version of it given in Ref.\,\cite{MP}.
(ii) next, the application of the procedure in Ref.\cite{CHENGETAL} to the
classical HOTD MP model would reduce it to a modified effective theory of
the same time derivative character as classical electrodynamics. The
procedure leads to field redefinitions plus additional contributions to the
interactions. (iii) finally, this resulting classical theory would be
considered as the correct starting point for quantization, which would be
carried along the standard lines. The resulting quantum theory would then
provide the basis for the calculation of interacting processes using the
perturbative scheme of quantum field theory (QFT). Some of these
steps have been already carried out in Refs.\cite{Reyes:2007qq}, for the
case of the scalar and fermion fields.

Perhaps we should emphasize at this stage that we are dealing with two
different classes of perturbations: the first one concerns only the LIV
parameters, occurs at the classical level and serves to define the correct
starting point for quantization. Once the resulting theory is quantized, the
usual QFT interacting processes can be calculated, corresponding to the
second class of perturbations. Both approximations should be made consistent
when predicting a result to a given order in any of the LIV parameters. In
this sense it is clear that we are not producing a quantum version of the
full MP model, but only one which is adapted to our basic requirement of
describing the LIV corrections as perturbations to QED.

Since the model respects observer (passive) Lorentz transformations we
consider the parameters $\rho ,\xi ,\eta _{1}$, $\eta _{2}$ and ${\bar M}$,
to be invariant under them. Nevertheless, the general form of the
four-vector describing the preferred frame is $n^\mu=\gamma(1, \mathbf{v})$,
with $1/\gamma=\sqrt{1-\mathbf{v}^2}$, so that highly boosted systems will
greatly amplify the values of the LIV parameters which are strongly
constrained in earth-based reference frames. Thus we also restrict the
observer Lorentz transformations to concordant reference frames which move
non-relativistically with respect to earth \cite{LEHNERT}. In the sequel we
will give a quantitative characterization of such allowed observers. A
further simplification is introduced by taking into account that the
parameters $\rho ,\xi ,\eta _{1}$ and $\eta _{2}$ are independent. In this
way we set the field $\phi$ together with the parameters
$\rho ,\eta_{1}$and $\eta _{2}$ equal to zero. Then we deal with a minimal
LIV extension of standard QED.

The paper is organized as follows. In section II we discuss the classical MP
modifications to electrodynamics. There we construct the corresponding
Hamiltonian formulation in terms of canonically transformed fields that
guarantee the appropriate normalization of the momentum squared terms in the
Hamiltonian density, that includes the identification of the interacting sector.
Section III deals with  the quantization of the model in
terms of standard creation-annihilation operators. The modified dispersion
relations are identified and the Hamiltonian is shown to be positive
definite for momenta $\mathbf{k}$ such that $|\mathbf{k}|<{\bar M}/(2 |\xi|)$.
In Section IV we construct the modified photon propagator in the Coulomb
gauge which is subsequently written in four dimensional notation by
incorporating the static Coulomb contribution appearing in the Hamiltonian.
Section V contains the physical motivation and specific proposal for our
prescription that allows to understand the quantum MP model as a tiny
perturbation of QED, according to the experimental and observational
evidence of highly suppressed LIV. A coarse graining mass scale $M<<{\bar M}$
is further introduced in the problem, dictated by the effective character of
the model, and signaling the onset of the modifications in the space-time
structure. In Section VI we set up the general structure  the electron self-energy
calculation including only the modified
photon propagator $(\xi \neq 0)$ interacting with standard fermions $%
(\eta_1=\eta_2=0)$. The scale $M$ is taken into account via a factor of the
Pauli-Villars type, designed to act as the appropriate regulator in the QED
limit. Also we perform a power expansion of the self-energy in terms of the
external momentum and identify those terms to be subjected to scrutiny
regarding their suppressed character and good QED limit in the next Section.
The general strategy for their calculation is presented in Section VII and
all the LIV contributions to order $\xi^2$ are accordingly obtained. One of
such calculations is presented in full detail, while we only write the
results for the remaining ones. In Section VIII we present a preliminary
study of the microcausality violation in the model by identifying the
space-like region where it occurs, together with an estimation of the
magnitude of such violation. The final Section IX contains a summary of the
work. The notation and conventions are stated in the Appendix \ref{APPA} which,
together with the Appendix \ref{APPB}, contain information relevant for the specific
calculations in the paper. In Appendix \ref{APPC} the relationship between the modified
photon propagator in different gauges  is stablished.  The last Appendix \ref{APPD}
includes the definitions of
the LIV contributions which calculation is not fully developed in the text.

\section{The model}

\label{secmodel}

With the simplifications stated above we consider the modified the photon
sector
\begin{equation}
\mathcal{L}_{\gamma}=-\frac{1}{4}F_{\mu \nu }F^{\mu \nu }+\frac{\xi }{2
\tilde{M}}\left( n^{\mu }F_{\mu \nu }\right) \left( n^{\alpha }\partial
_{\alpha }\right) \left( n_{\rho
}\epsilon^{\rho\nu\kappa\lambda}F_{\kappa\lambda}\right) -J^{\mu }A_{\nu },
\end{equation}%
where the electromagnetic current $J^{\mu }$ will be subsequently realized
in terms of unmodified spin $1/2$ fermions, according to the choice $%
\eta_1=\eta_2=0$. Our general strategy will be first to  quantize
the photons and subsequently to consider the interaction, via the standard
minimal coupling, with the unmodified quantum fermions.

The equations of motion in the Lorentz gauge are

\begin{equation}
\left( \eta ^{\nu \lambda }\partial ^{2}-2g\left( n\cdot \partial \right)
^{2}n_{\rho }\epsilon ^{\rho \nu \sigma \lambda }\partial _{\sigma }\right)
A_{\lambda }=J^{\nu }.  \label{EQMOTLG}
\end{equation}

In order to get a better control of the LIV modifications we find it
convenient to work in the Hamiltonian scheme, so that we switch to a $3+1$
canonical formulation of the problem. Taking advantage of the remaining
observer Lorentz invariance of the model, we choose to work in the rest
frame $n^{\mu }=(1,\mathbf{0})$, where the free modified photon contribution
is
\begin{equation}
\mathcal{L}_{\gamma }=\frac{1}{2}\left( \dot{A}^{i}+\partial
_{i}A^{0}\right) ^{2}-\frac{1}{4}F_{ij}F^{ij}+g\epsilon ^{ijk}\dot{A}%
^{i}\partial_{j}\dot{A}^{k}-J^{\mu }A_{\mu },\;\;\;g\mathbf{=}\frac{\xi }{%
\tilde{M}} .  \label{LREST}
\end{equation}
This choice has the advantage that, up to a total derivative, the resulting
system is not of the HOTD type. Nevertheless, it exhibits in a simpler
setting most of the questions associated to the quantization of the full MP
model. In addition, let us emphasize that we will carry the quantization
without any approximation in the parameter $g$.

The canonical approach gives the following momenta
\begin{equation}
\Pi _{0}=\frac{\partial \mathcal{L}_{\gamma }}{\partial \dot{A}_{0}}%
=0,\;\;\;\Pi _{i}=\frac{\partial \mathcal{L}_{\gamma }}{\partial \dot{A}^{i}}%
=\dot{A}^{i}+\partial _{i}A^{0}+2g\epsilon ^{ijk}\partial _{j}\dot{A}^{k},
\label{MOMENTA}
\end{equation}%
together with their Poisson brackets,
\begin{equation}
\left\{ A^{i}(t,\mathbf{x}),\Pi _{j}(t,\mathbf{y})\right\} =\delta
_{j}^{i}\delta ^{3}(\mathbf{x}-\mathbf{y}).
\end{equation}%
The next step is to construct the Hamiltonian density $\mathcal{H}_{C}=\Pi
_{i}\dot{A}^{i}-\mathcal{L}$, which is
\begin{equation}
\mathcal{H}_{C}=\Pi _{i}\dot{A}^{i}-\frac{1}{2}\left( \dot{A}^{i}+\partial
_{i}A^{0}\right) ^{2}+\frac{1}{4}F_{ij}F^{ij}-g\epsilon ^{ijk}\dot{A}%
^{i}\partial _{j}\dot{A}^{k}+J^{\mu }A_{\mu }.  \label{HAMC}
\end{equation}%
In order to write the velocities in terms of the momenta it is convenient to
consider the combination $\Pi _{i}-\partial _{i}A^{0}$ together with the
operator
\begin{equation}
M^{ik}=\left( \delta ^{ik}+2g\epsilon ^{ijk}\partial _{j}\right),
\label{OPB}
\end{equation}%
in the second Eq.(\ref{MOMENTA}). To solve for the velocities we need the
inverse of the operator $M^{ik}$ for which we obtain the exact non-local
expression
\begin{equation}
\left( M^{-1}\right)^{ij}=\frac{1}{\left( 1+4g^{2}\mathbf{\nabla }%
^{2}\right) }\left(\delta ^{ij}-2g\epsilon ^{irj}\partial
_{r}+4g^{2}\partial_{i}\partial _{j}\right) .  \label{INVOPB}
\end{equation}%
In this way we solve
\begin{equation}
\dot{A}^{i}=\left( M^{-1}\right) ^{ij}\left( \Pi _{j}-\partial
_{j}A^{0}\right) ,  \label{VEL}
\end{equation}%
which we substitute in Eq. (\ref{HAMC}). The result is
\begin{equation}
\mathcal{H}_{C}=\left( \Pi _{p}-\partial _{p}A^{0}\right) \frac{1}{2}\left(
M^{-1}\right)^{pr}\left( \Pi_{r}-\partial_{r}A^{0}\right) -\frac{1}{2}\left(
\partial_{i}A^{0}\right)^{2}+\frac{1}{4}F_{ij}F^{ij}+J^{\mu }A_{\mu }.
\label{HAMCFIN}
\end{equation}%
Integrating by parts and using some of the properties for $\left(
M^{-1}\right) ^{ij}$ written in the Appendix A, we arrive at
\begin{equation}
\mathcal{H}_{C}=\frac{1}{2}\Pi _{p}\left( M^{-1}\right) ^{pr}\Pi _{r}+\left(
\partial _{p}\Pi _{p}+J^{0}\right) A^{0}+\frac{1}{4}F_{ij}F^{ij}-J^{i}A^{i}.
\label{HAMCFIN1}
\end{equation}%
It can be verified that the corresponding Hamilton equation of motion
reproduces the correct expression (\ref{VEL}) for $\dot{A}^{i}.$

The  canonical  variables can be written in the
convenient form
\begin{equation}
\Pi _{i}^{T} =\dot{A}_{T}^{i}+2g\epsilon ^{ijk}\partial _{j}\dot{A}%
_{T}^{k}, \qquad
\Pi _{i}^{L}=\dot{A}_{L}^{i}+\partial _{i}A^{0}, \qquad
A_{L}^{i}=\frac{1}{\nabla ^{2}}\partial _{i}(\partial _{k}A^{k}),  \label{AL}
\end{equation}%
where we are using the standard definition for a transverse ($T$) and
longitudinal ($L$) decomposition of a vector field $\mathbf{U}=\mathbf{U}%
_{T}+\mathbf{U}_{L}$, where $\mathbf{\nabla }\cdot \mathbf{U}_{T}=0,\;\;%
\mathbf{\nabla }\times \mathbf{U}_{L}=0$. In the case of the velocities the separation leads to
\begin{equation}
\dot{A}_{T}^{i} =\frac{1}{W^2 }%
\left( \delta ^{ij}-2g\epsilon ^{irj}\partial _{r}\right) \Pi^T_{j},
\qquad
\dot{A}_{L}^{i}=\frac{1}{W^2 }\Pi^L
_{i}+\partial _{i}\left( \frac{4g^{2}}{W^2 }\partial _{j}\Pi^L _{j}-A^{0}\right),  \label{LONG}
\end{equation}
with the notation\ $W=\sqrt{ 1+4g^{2}\nabla ^{2}}$.
As in the usual case, $A_{0}$ is a Lagrange multiplier leading to the Gauss
law as a secondary constraint
\begin{equation}
\Theta =\partial _{i}\Pi _{i}+J^{0}=0,  \label{GAUSS}
\end{equation}%
which can also be understood as arising from the time derivative $\dot{\Pi}%
_{0}$ of the primary constraint $\Pi _{0} \approx 0$. The evolution $\dot{%
\Theta}\approx 0$ leads to current conservation in such a way that we have
only two first class constraints as in the standard case.
In terms of transverse and longitudinal variables the Gauss law is written as%
\begin{equation}
\partial _{i}\Pi _{i}^{L}+J^{0}=0.  \label{GAUSSLONG}
\end{equation}%
The equation of motion
\begin{equation*}
\partial ^{i}F_{i0}=+J^{0},
\end{equation*}%
yields
\begin{equation}
A^{0}=-\frac{1}{\nabla ^{2}}\left( J^{0}+\partial _{0}\partial
_{i}A^{i}\right) .
\end{equation}%
At this stage we select the Coulomb gauge by choosing
\begin{equation}
A^{0}=-\frac{1}{\nabla ^{2}}J^{0},\quad \Pi _{0}=0,  \qquad
\partial _{k}A^{k} =0\quad \rightarrow \quad A_{L}^{i}(x)=0, \qquad \Pi _{i}^{L}
=-\frac{1}{\nabla ^{2}}\partial _{i}J^{0}. \label{COULOMB}
\end{equation}%
The dynamical variables are contained only in the transverse modes
\begin{equation}
\Pi _{i}^{T}=\dot{A}_{T}^{i}+2g\epsilon ^{ijk}\partial _{j}\dot{A}%
_{T}^{k},\qquad  A_{T}^{i}(x),  \label{ATPIT}
\end{equation}%
which satisfy the Dirac brackets%
\begin{equation}
\left\{ A_{T}^{i}(t,{\bf x}),\;\Pi _{m}^{T}(t,{\bf y})\right\} =\left( \delta ^{im}-\frac{%
\partial _{x}^{i}\partial _{x}^{m}}{\nabla _{x}^{2}}\right) \delta ^{3}({\bf x}-{\bf y}).
\label{PPT}
\end{equation}%
Using repeated integration by parts in the Hamiltonian, together with the
transversality condition, we arrive at
\begin{equation}
H_{C}=\int d^{3}x\;\left( \frac{1}{2}\Pi _{p}^{T}\left( M^{-1}\right)
^{pr}\Pi _{r}^{T}+\frac{1}{2}J^{0}\left( -\frac{1}{\nabla ^{2}}\right) J^{0}+%
\frac{1}{4}F_{ij}F^{ij}-J^{i}A_{T}^{i}\right) .  \label{HAMCFIN3}
\end{equation}

Our final goal is to express the dynamical fields in terms of
creation-annihilation operators, corresponding to modified frequency modes,
satisfying standard bosonic commutation relations that will reproduce the
field commutation relations arising from the correspondence principle
applied to the respective Dirac brackets. To this end it is necessary that
the relation $\Pi =\dot{A}$ \ holds, which is equivalent to require that the
kinetic term of the Hamiltonian density be normalized as $\frac{1}{2}\Pi ^{2}
$. In order to achieve this we perform the canonical transformation ($%
\mathbf{A}_{T}\rightarrow \mathbf{\bar{A}}_{T},\;\mathbf{\Pi }%
^{T}\rightarrow \mathbf{\bar{\Pi}}^{T}\;$) given by
\begin{equation}
A_{T}^{i}=\frac{\sqrt{1+W}}{\sqrt{2}W}\left[ \delta ^{iq}-\frac{2g}{\left(
1+W\right) }\epsilon ^{imq}\partial _{m}\right] ,\;\;\;\;\;\;%
\;\Pi _{r}^{T}=\frac{\sqrt{1+W}}{\sqrt{2}}\left[ \delta ^{rq}+\frac{2g}{%
\left( 1+W\right) }\epsilon ^{rmq}\partial _{m}\right] \bar{\Pi}_{q}^{T}.
\label{TCA}
\end{equation}%
The non-zero transverse Dirac brackets for the  variables $\bar{A}_{T}^{i}$ and $ \bar{\Pi}_{j}^{T}$
are the same as
in Eq. (\ref{PPT}) in virtue of the canonical character of the transformation.
Rewriting the Hamiltonian (\ref{HAMCFIN3}) in terms of the new variables
leads to
\begin{equation}
H_{C}=\int d^{3}x\;\Bigg( \frac{1}{2}\bar{\Pi}_{p}^{T}\bar{\Pi}_{p}^{T}+%
\frac{1}{2}\bar{A}_{T}^{r}\left(-\frac{
\nabla ^{2}}{ W^2 }\right) \left[ \delta ^{rp}-2g\epsilon ^{rnp}\partial _{n}\right]
\bar{A}_{T}^{p}+\frac{1}{2}J^{0}\left( -\frac{1}{\nabla ^{2}}\right)
J^{0}-J^{i}A_{T}^{i}(\bar{A}_{T})\Bigg) .  \label{HCANFIN4}
\end{equation}
Let us emphasize that in the last interaction term $A_{T}^{i}$ is a
functional of the dynamical field $\bar{A}_{T}^j$. In this sense the
electromagnetic vertex will be modified with respect to the latter
field but will retain the usual structure with respect to the
former. In this way, some care is required when implementing the
perturbation theory starting from the zeroth order Hamiltonian
written in terms of ${\bar A}_T^i$ and ${\bar \Pi}^T_j$.

\section{The Quantum Theory}

Now we have the basic ingredients to proceed with the quantization of the
modified photon field. We start form the usual plane wave expansion of the
operator $\bar{A}_{T}^{i}(x)$
\begin{equation}
\bar{A}_{T}^{i}(x)=\int \frac{d^{3}\mathbf{k}}{\sqrt{(2\pi )^{3}}}%
\,\sum_{ \lambda= \pm 1 }\sqrt{\frac{1}{2\omega _{\lambda }(\mathbf{k)}}}\left[
\,a_{\lambda }(\mathbf{k})\,\varepsilon ^{i}(\lambda ,\mathbf{k}%
)e^{-ik(\lambda )\cdot x}+a_{\lambda }^{\dag }(\mathbf{k})\,\varepsilon
^{i\ast }(\lambda ,\mathbf{k})e^{+ik(\lambda )\cdot x}\right] ,
\label{AMUBAREXP}
\end{equation}%
in terms of creation-annihilation operators $a_{\lambda }^{\dag }(\mathbf{k}%
),\; a_{\lambda }(\mathbf{k})$, respectively. The notation is
\begin{equation}
\left[ k(\lambda )\right] _{\mu }=(\omega _{\lambda }(\mathbf{k)},-\mathbf{k}%
),\;\;\;\;k(\lambda )\cdot x=\omega _{\lambda }(\mathbf{k)}x^{0}-\mathbf{%
k\cdot x},
\end{equation}%
where the modified normal frequencies will be consistently determined.
The properties of the polarization vectors $\varepsilon^{i}(\lambda ,\mathbf{%
k}), \, \lambda=\pm 1$, chosen in the circularly polarized (helicity) basis, are collected in
the Appendix B. The momenta are given by
\begin{equation}
\bar{\Pi}_{i}^{T}(x)=\int \frac{d^{3}\mathbf{k}}{\sqrt{(2\pi )^{3}}}%
\,\sum_{\pm \lambda }\sqrt{\frac{1}{2\omega _{\lambda }(\mathbf{k)}}}\left[
\,\left( -i\omega _{\lambda }\right) a_{\lambda }(\mathbf{k})\,\varepsilon
^{i}(\lambda ,\mathbf{k})e^{-ik(\lambda )\cdot x}+\left( i\omega _{\lambda
}\right) a_{\lambda }^{\dag }(\mathbf{k})\,\varepsilon ^{i\ast }(\lambda ,%
\mathbf{k})e^{+ik(\lambda )\cdot x}\right] .  \label{PIMUBAREXP}
\end{equation}
Assuming the standard creation-annihilation commutation rules
\begin{equation}
\left[ a_{\lambda }(\mathbf{k}),\;a_{\lambda ^{\prime }}^{\dag }(\mathbf{k}%
^{\prime })\right] =\delta _{\lambda \lambda ^{\prime }}\delta ^{3}(\mathbf{%
k-k}^{\prime })  \label{CAOPCR}
\end{equation}%
and starting from (\ref{AMUBAREXP}) and (\ref{PIMUBAREXP})\ we recover the
basic field commutator at equal times
\begin{equation}
\,\left[ \bar{A}_{T}^{i}(t, {\bf x}),\bar{\Pi}_{j}^{T}(t, {\bf y})\right] =i\left( \delta
^{ij}-\frac{\partial _{xi}\partial _{xj}}{\nabla ^{2}}\right) \delta
^{3}\left( \mathbf{{\bf x}-{\bf y}}\right) ,
\end{equation}%
which is the expected result after the canonical transformation.
The corresponding equations of motion are
\begin{equation}
\bar{\Pi}_{q}^{T}=\partial _{0}\bar{A}_{T}^{q},\qquad
\dot{\Pi}_{r}^{T}= \frac{ \nabla^{2}}{W^2}  \left[ \delta ^{rp}-2g\epsilon ^{rnp}\partial _{n}\right]
\bar{A}%
_{T}^{p}.
\end{equation}%
Going to the momentum space we can obtain the modified dispersion relations
from
\begin{equation}
\omega ^{2}\mathbf{\bar{A}}_{T}(k)= \frac{|\mathbf{k|}^{2}}{ 1-4g^{2}|%
\mathbf{k|}^{2} } \left[ \mathbf{\bar{A}}_{T}(k)-2gi\mathbf{k}\times \mathbf{%
\bar{A}}_{T}(k)\right] ,
\end{equation}%
which reduces to the diagonalization%
\begin{equation}
i\mathbf{k\times \;\rightarrow \;\;}\lambda |\mathbf{k|,\;\;\;\;\mathbf{A}%
_{T}\rightarrow \mathbf{A}_{T}^{\lambda },\;\;}
\end{equation}%
when the vector potential is expressed in the helicity basis. In this way
\begin{equation}
\omega _{\lambda }^{2}\mathbf{\bar{A}}^{\lambda}_T=\left( \frac{|\mathbf{k|}%
^{2}}{ 1-4g^{2}|\mathbf{k|}^{2} }\right) \left[ \mathbf{1}-2\lambda g|%
\mathbf{k|}\right] \mathbf{\bar{A}}^{\lambda}_T,
\end{equation}%
yielding the modified energy-momentum relation
\begin{equation}
\omega _{\lambda }^{2}\left( \mathbf{k}\right) =\frac{|\mathbf{k|}^{2}}{%
\left[ 1+2\lambda g|\mathbf{k|}\right] },  \label{RELDISFIN}
\end{equation}%
which is exact in $g$. With no loss of generality we assume from now on that
$g>0$. Let us notice that the four-vector $[k(\lambda=+1)]_\mu$ is
spacelike, while $[k(\lambda=-1)]_\mu$ is timelike. At this stage we are
confronted with two problems that arise rather frequently in LIV theories:
(i) on one hand, the frequency $\omega_-({\mathbf k})$ will become imaginary when $|%
\mathbf{k}|> 1/(2g)$ and diverges when $|\mathbf{k}|=|\mathbf{k}|_{max}=
1/(2g)$. From an intuitive point of view we consider $1/(2g)$ as the
analogous of the value $|\mathbf{k}|_{\max }=\infty$ in the standard case
and we will cut all momentum integrals at this value. The introduction of
the coarse graining scale $M<<{\bar M}$, explained in more detail in Section
V, effectively produces the more stringent and smooth cut-off
\begin{equation}
g|\mathbf{k}|< gM << 1.  \label{CONDITION}
\end{equation}
(ii) on the other hand, since $[k(\lambda=+1)]_\mu$ is spacelike, we can
always perform an observer Lorentz transformation such that $\omega_+({\mathbf k})$
becomes negative thus introducing stability problems in the model. For a
given momentum $\mathbf{k}$ this occurs for $1/\sqrt{1+2g|\mathbf{k}|}<|%
\mathbf{v}| <1$. Then, the condition (\ref{CONDITION}) leads to the
requirement that the allowed concordant frames in which the quantization
will remain consistent are such that $\gamma < 1/\sqrt{2gM}$, with respect
to the rest frame.

Our next step is to verify that the resulting free ($J^{\mu }=0$)
Hamiltonian is in fact positive definite and has the expected expression in
terms of the previously introduced creation-annihilation operators.
Let us begin with the kinetic term%
\begin{equation}
H_{KE}=\frac{1}{2}\int d^{3}x\,\bar{\Pi}_{i}^{T}\bar{\Pi}_{i}^{T},
\end{equation}%
which leads to
\begin{equation}
H_{KE}=\frac{1}{2}\int \,d^{3}\mathbf{k}\sum_{\lambda }\left[ \left( -\frac{%
\omega _{\lambda }(\mathbf{k)}}{{2}}\right) a_{\lambda }(\mathbf{k}%
)a_{\lambda }(-\mathbf{k})\,e^{-i2\omega _{\lambda }(\mathbf{k})t}+\frac{%
\omega_{\lambda }(\mathbf k)}{{2}}a_{\lambda }^{\dag }(\mathbf{k})\,a_{\lambda }(%
\mathbf{k})+h.c.\right],
\end{equation}%
in terms of the creation-annihilation operators.

The potential term contribution is
\begin{equation}
H_{POT}=\frac{1}{2}\int d^{3}x\bar{A}_{T}^{r}\left(-\frac{\nabla ^{2}}{W^2 }\right)\,
\left[ \delta ^{rp}-2g\epsilon
^{rnp}\partial _{n}\right] \bar{A}_{T}^{p},
\end{equation}%
which analogously reduces to
\begin{equation}
H_{POT}=\frac{1}{2}\int d^{3}\mathbf{k}\,\,\sum_{\pm \lambda }\;\;\left[
\left( \frac{\omega _{\lambda }(\mathbf{k)}}{2}\right) \left[ a_{\lambda }(%
\mathbf{k})\,a_{\lambda }(-\mathbf{k})\,\right] e^{-i2\omega _{\lambda }(%
\mathbf{k})t}+\frac{\omega _{\lambda }(\mathbf{k)}}{2}a_{\lambda }^{\dag }(%
\mathbf{k})\,a_{\lambda }(\mathbf{k})+h.c.\right].
\end{equation}
Here we have made use of the dispersion relations (\ref{RELDISFIN}),
together with Eqs. (\ref{ORT1}), (\ref{ORT2}). This leads to the expected
final expression
\begin{equation}
H=\frac{1}{2}\int d^{3}\mathbf{k}\,\sum_{\lambda }\left[ \,a_{\lambda }(%
\mathbf{k})a_{\lambda }^{\dag }(\mathbf{k})\,+a_{\lambda }^{\dag }(\mathbf{k}%
)a_{\lambda }(\mathbf{k})\right] \omega _{\lambda }(\mathbf{k)},
\end{equation}%
arising from the cancelation of the time dependent terms and
including the modified frequencies (\ref{RELDISFIN}). Thus the
Hamiltonian is hermitian as far as the frequencies remain real,
which is the case in the region $|{\mathbf k}|< 1/(2g)$.

\section{The Photon Propagator}

In this section we calculate the free modified photon propagator starting
from the definition\thinspace
\begin{eqnarray}
i\bar{\Delta}_{ij}(x,y) &\equiv &\left\langle 0\right| T\left( {\bar{A}%
_{i}^{T}\left( x\right) \bar{A}_{j}^{T}\left( y\right) }\right) \left|
0\right\rangle \hfill  \notag \\
&=&\theta\left( {x^{0}-y^{0}}\right) \left\langle 0\right| {\bar{A}}%
_{i}^{T}\left( x\right) {\bar{A}}_{j}^{T}\left( y\right) \left|
0\right\rangle +\theta\left( {y^{0}-x^{0}}\right) \left\langle 0\right|
\bar{A}_{j}^{T}\left( y\right) \bar{A}_{i}^{T}\left( x\right) \left|
0\right\rangle ,  \label{phot-propg}
\end{eqnarray}%
where $\bar{\Delta}_{ij}(x,y)=\bar{\Delta}_{ij}(x-y)$ as can be seen from
the expression
\begin{equation}
\left\langle 0\right| \bar{A}_{i}^{T}\left( x\right) \bar{A}_{j}^{T}\left(
y\right) \left| 0\right\rangle =\int {\frac{{d^{3}}\mathbf{k}}{{\left( {2\pi
}\right) ^{3}}}}\sum\limits_{\lambda }{\frac{1}{{2\omega _{\lambda }(\mathbf{%
k})}}}\mathrm{\ }e^{-i(x-y) \cdot \, k(\lambda )}\, \varepsilon
_{i}( \lambda, \mathbf{\hat{k}})\,\, \varepsilon _{j}^{\ast }(
\lambda, \mathbf{\hat{k}}).
\end{equation}
Here we introduce the notation%
\begin{equation}
F_{ij}(\lambda ,\mathbf{\hat{k}})=\varepsilon _{i}(\lambda, \mathbf{%
\hat{k}})\,\, \varepsilon _{j}^{\ast }(\lambda, \mathbf{\hat{k}}%
),
\end{equation}%
which leads to the second vacuum expectation value in (\ref{phot-propg})
\begin{equation}
\left\langle 0\right| \bar{A}_{j}^{T}\left( y\right) \bar{A}_{i}^{T}\left(
x\right) \left| 0\right\rangle =\int {\frac{{d^{3}}\mathbf{k}}{{\left( {2\pi
}\right) ^{3}}}}\sum\limits_{\lambda }{\frac{1}{{2\omega _{\lambda }(\mathbf{%
k})}}}\mathrm{\ }e^{i\left( {x-y}\right) \cdot k(\lambda )}F_{ji}(\lambda ,%
\mathbf{\hat{k}}).
\end{equation}%
We are interested in expressing the propagator
\begin{equation}
\bar{\Delta}_{ij}(z)=\int \frac{d^{4}k}{\left( 2\pi \right) ^{4}}%
e^{-ik\cdot z }\Delta _{ij}(k),
\end{equation}%
with $z^{\mu
}=x^{\mu }-y^{\mu }$, in momentum space. To this end we start from the expression
\begin{equation}
i\bar{\Delta}_{ij}(z) = \int \frac{d^{3}k}{(2\pi )^{3}}e^{+i\mathbf{k\cdot z%
}}\left[ \theta(z_{0})\sum_{\pm \lambda }e^{-i{\omega_{\lambda }}z_{0}}\frac{%
1}{{2\omega _{\lambda }(\mathbf{k})}}F_{ij}(\lambda ,\mathbf{\hat{k}})
+\theta(-z_{0})\sum_{\pm \lambda }e^{i{\omega_{\lambda }}z_{0}}%
\frac{1}{{2\omega _{\lambda }(\mathbf{k})}}F_{ji}(\lambda ,-\mathbf{\hat{k}})%
\right],
\end{equation}%
and introduce the standard representation
\begin{equation}
\theta(z_{0})=i\lim_{\epsilon \rightarrow 0}\int_{-\infty }^{\infty }\frac{%
d\alpha }{2\pi }\frac{e^{-i\alpha z_{0}}}{\alpha +i\epsilon },
\end{equation}%
in order to calculate the corresponding Fourier transform. The result is
\begin{equation}
\bar{\Delta}_{ij}(k)=\sum\limits_{\lambda }\left( \frac{
\varepsilon_{i}(\lambda, \mathbf{\hat{k}})\,\, \varepsilon_{j}^{\ast }(
\lambda, \mathbf{\hat{k})}}{2\omega_{\lambda }\left( {%
k_{0}-\omega_{\lambda }+i\varepsilon }\right)}
 -\frac{\varepsilon_{j}(\lambda, -\mathbf{\hat{k}})\,\,
\varepsilon_{i}^{\ast}(\lambda, -\mathbf{\hat{k}})}{2\omega_{\lambda }
\left( k_{0}+\omega_{\lambda }-i\varepsilon\right)} \right). \label{PROPCERO}
\end{equation}
Using Eqs. (\ref{EIJ1}) and (\ref{EIJ2}) we rewrite the propagator in the
form
\begin{equation}
\bar{\Delta}_{ij}\left( k\right) =\frac{1}{2}{\sum\limits_{\lambda }{\frac{{1%
}}{{\left( {k_{0}^{2}-}\left( {\omega_{\lambda }-i\varepsilon }\right)
^{2}\right) }}}}\left( \left[ \delta _{ij}-\frac{k_{i}k_{j}}{|\mathbf{k|}^{2}%
}\right] +i\lambda \left[ \epsilon ^{ijm}\frac{k_{m}}{|\mathbf{k|}}\right]
\right).  \label{PROP1}
\end{equation}
After performing the summations according to (\ref{S1}), (\ref{S2}) and (\ref{S3}) we
arrive at the following expression for the modified photon propagator in the
Coulomb gauge
\begin{equation}
\bar{\Delta}_{ij}\left( k\right) =\frac{1}{\left(\left( k{^{2}}\right)
^{2}-4g^{2}|\mathbf{k|}^{2}k{_{0}^{4}}\right)}\left[ \left( k{^{2}}-4g^{2}|%
\mathbf{k|}^{2}k{_{0}^{2}}\right) \left[ \delta _{ij}-\frac{k_{i}k_{j}}{|%
\mathbf{k|}^{2}}\right] -2g|\mathbf{k|}^{2}\epsilon ^{ijm}ik_{m}\right].
\label{PROP1EXPL}
\end{equation}
Let us verify the correct limits when $g=0$, where $\omega ^{+}=\omega
^{-}=\omega =|\mathbf{k|}$. In this case the first sum in the RHS of Eq.(\ref%
{PROP1}) gives the standard transverse propagator, while the second sum
cancels out.

We would like now to extend the above propagator, which is defined in the
transverse sector, to the whole four dimensional space in such a way that
the current-current interaction is described by
\begin{equation}
\frac{1}{2}\int d^{4}k\;J^{\mu }(-k){\bar \Delta}_{\mu \nu }(k)J^{\nu }(k).
\end{equation}%
This is achieved by incorporating in Eq. (\ref{PROP1}) the Coulomb term
appearing in (\ref{HCANFIN4}) in a manner analogous to that described in
Ref. \cite{Weinbergbook}. The final result is
\begin{equation}
\bar{\Delta}_{\mu \nu }=-\frac{1}{2}\sum_{\lambda }\frac{{1}}{{\left( k{%
_{0}^{2}-{\omega _{\lambda }}^{2}+i\epsilon }\right) }}\left( \eta _{\mu \nu
}+\left( \frac{{\omega _{\lambda }}^{2}}{|\mathbf{k|}^{2}}-1\right) \delta
_{\mu 0}\delta _{\nu 0}-i\lambda n_{\rho }\epsilon _{\mu \nu }^{\;\;\;\rho
\sigma }\frac{k_{\sigma }}{|\mathbf{k|}}\right) ,  \label{FINPROP0}
\end{equation}%
where we have reinserted the vector $n^{\rho }=(1,\mathbf{0})$.

The last step in the construction is to perform the sums over $\lambda $ in (%
\ref{FINPROP0}) using the corresponding expressions in the Appendix B. The
result is
\begin{equation}
\bar{\Delta}_{\mu \nu }=-\frac{1}{\left( \left( k{^{2}}\right) ^{2}-4g^{2}|%
\mathbf{k|}^{2}k{_{0}^{4}}\right) }\left[ \eta _{\mu \nu }\left( k{^{2}}%
-4g^{2}|\mathbf{k|}^{2}k{_{0}^{2}}\right) +4g^{2}|\mathbf{k|}^{2}k{_{0}^{2}}%
\delta _{\mu 0}\delta _{\nu 0}+2gn_{\rho }\epsilon _{\mu \nu }^{\;\;\;\rho
\sigma }\left( ik_{\sigma }\right) |\mathbf{k|}^{2}\right] .  \label{PROPFIN}
\end{equation}

The propagator obtained directly from the equations of motion (\ref{EQMOTLG}%
) in the Lorentz gauge is
\begin{equation}
\Delta _{\mu \nu }(k)=\frac{1}{((k^{2})^{2}-4g^{2}\left| \mathbf{k}%
\right|^{2}k_{0}^{4})}\left[ -k^{2}\eta _{\mu \nu }+2igk_{0}^{2}\epsilon
^{lmr}k_{m}\eta _{l\mu }\eta _{r\nu }-\frac{4g^{2}k_{0}^{4}}{k^{2}}%
k_{l}k_{r}\delta _{\mu }^{l}\delta _{\nu }^{r}+\frac{4g^{2}k_{0}^{4}{\left|
\mathbf{k}\right| }^{2}}{k^{2}}\eta _{0\mu }\eta _{0\nu }\right] .
\label{LORGAUGE}
\end{equation}
In Appendix \ref{APPC} we have calculated  the propagator $\Delta_{ij}$
corresponding to the fields $A_T^i$ starting from ${\bar \Delta}_{ij}$
given by (\ref{PROP1EXPL}) and performing the canonical transformation (\ref{TCA}).
Moreover, the subsequent inclusion of the Coulomb term in $\Delta_{ij}$ leads exactly
to the four dimensional propagator $\Delta_{\mu\nu}$ in (\ref{LORGAUGE}). It is
important to emphasize that the Hamiltonian (\ref{HCANFIN4}) has a noninteracting
sector described by the fields ${\bar A}_T^i, {\bar \Pi}^T_j$ but induces an
interaction density given by $ J_iA_T^i \rightarrow N[J_\mu A^\mu]$, where $A^\mu$
propagates according to (\ref{LORGAUGE}).

\section{The prescription defining the effective quantum model}

The main goal of this work is to study the possibility of defining the MP
model as a perturbative extension of standard QED, that is to say as a model
which continuously interpolates between a LIV theory and a Lorentz
preserving one. This is to a large extent motivated by the very stringent
experimental and observational limits set upon the parameters that codify
such LIV. A construction exhibiting this interpolating characteristic has
been already presented in Ref. \cite{ALFARO},\ but there the LIV was
codified by a dimensionless parameter, as opposed to the situation here.
As we will explain the sequel, the effective character of the model requires
the introduction of an additional mass scale $M$ that provides the
analogous dimensionless parameter $(gM)$.

Another point that requires attention is the upper limit $|\mathbf{k}|_{\max
}=1/(2g)$ set by the modified dispersion relations (\ref{RELDISFIN}), which
guarantees the absence of imaginary frequencies together with that of a
non-hermitian Hamiltonian. We consider these facts as indications of the
effective character of the model. Assuming for a moment that $1/g\approx
E_{QG}\approx M_{Planck}\;$the above upper limit would mean that one is
probing distances of the order of the Planck length, where we expect quantum
gravity effects to be so important that the continuum properties of space
might be no longer valid, thus invalidating the use of an standard effective
field theory. This means that we need to introduce an additional
coarse-graining scale $M$ under which we can safely consider space as a
continuum and apply effective field theory methods. Thus we require
\begin{equation}
M << \frac{1}{g}.  \label{RELSCALES}
\end{equation}
In this way, the upper limit $|\mathbf{k}|_{\max }=1/(2g)$ can be considered
as a mathematical limitation in our model, analogous to $|\mathbf{k}|_{\max
}=\infty$ in the standard case. The physical limitation of the model is
settled by the scale $M$ and requires to be imposed by an adequate smooth
regularization procedure that cuts down the corresponding degrees of freedom
over this scale, which occurs a long way before energies of the order $%
\approx 1/g$ are reached. In this manner the relation (\ref{RELSCALES}) imposes a
definite prescription to recover standard QED: (i) first set $g\rightarrow 0$
 for fixed $M$ and (ii) then set $M\rightarrow \infty$. Let us emphasize that at the level
of the effective model, the theory is finite and certainly will have an
explicit dependence upon the physical parameters $g$ and $M$. Now comes the
question on how do we introduce the scale $M$. Intuitively we think of $M$
as the parameter that will regularize the divergent integrals that will
appear in the limit$\;g\rightarrow 0$ describing standard QED. This suggests
that we introduce this parameter via a Lorentz covariant smooth function $%
\mathcal{I}(k)$, of the Pauli-Villars type for example, with the
same characteristics that one would require in order to regulate
standard QED. A natural choice for $\mathcal{I}(k)$ in our
calculation of the electron self-energy is
\begin{equation}
\frac{1}{k^{2}-m^{2}+i\epsilon }\mathcal{I}(k) = \frac{1}{%
k^{2}-m^{2}+i\epsilon }-\frac{1}{k^{2}-M^{2}+i\epsilon }
= \frac{1}{k^{2}-m^{2}+i\epsilon }\,\left(\frac{M^{2}}{M^{2}-k^{2}-i%
\epsilon }\right),\qquad M>>m.
\end{equation}%
In this way we are also imposing no additional LIV besides that arising from
the original modifications to the dynamics encoded in the parameter $g$.

\section{The electron self-energy}

\label{secselfenergy}

As a first step in testing the proposed construction we consider the
calculation of the electron self-energy with the dynamical modifications
introduced only via the LIV photon propagator. Let us recall that the
perturbation theory based upon the Hamiltonian (\ref{HCANFIN4}) indicates
that the photon propagates with $\Delta_{\mu\nu}$ given by (\ref{LORGAUGE}).
Moreover, we will focus upon the LIV contributions that could produce fine-tuning
problems associated to the would be divergent contributions arising in the
limit $g\rightarrow 0$.

The starting point is
\begin{equation}
\Sigma ^{g}(p)=-ie^{2}\int \frac{d^{4}k}{(2\pi )^{4}}\gamma ^{\mu }\left[
\frac{(\gamma \left( p-k\right) +m)}{((p-k)^{2}-m^{2}+i\epsilon )}\right]
\gamma ^{\nu }{\Delta }_{\mu \nu }(k)\,\mathcal{I}(k)\,\,\theta \left( \frac{%
1}{2g}-|\mathbf{k}{}|\right),   \label{SELFENERGY}
\end{equation}%
where we have introduced the scale $M$ via
\begin{equation}
\mathcal{I}(k)=\frac{M^{2}}{M^{2}-k^{2}}.
\end{equation}%
The $\theta $-function is there to guarantee the reality of the frequencies $%
\omega _{\lambda }(|\mathbf{k}{}|)$ entering the calculation of the photon
propagator in Section IV. Let us observe that the expression (\ref%
{SELFENERGY}) is finite.

Next we find it convenient to expand the self energy in powers of the
external momentum%
\begin{equation}
\Sigma ^{g}(p)=\Sigma _{p=0}^{g}+\left( \frac{\partial \Sigma ^{g}}{\partial
p^{\mu }}\right) _{p=0}p^{\mu }+\frac{1}{2}\left( \frac{\partial \Sigma ^{g}%
}{\partial p^{\mu }\partial p^{\nu }}\right) _{p=0}p^{\mu }p^{\nu }+O(p^{3}),
\label{SIGMAEXP}
\end{equation}%
where each coefficient in the expansion \ is a matrix written in terms of
some elements of the basis in the $4\times 4$ space of the Dirac matrices.
We have considered up to second derivatives in the external momentum because the
additional corrections to the numerator of the photon propagator (\ref{LORGAUGE})
of  order $gk$ and  $(gk)^2$ make those derivatives power counting  divergent,
as opposed to the QED case.
The fact that we are violating Lorentz transformations in the boost
sector, while maintaining rotational invariance would naturally split the
above expansion into a time plus space structure. The expansion of the above
coefficients in the gamma matrix basis will be denoted by
\begin{equation}
\Sigma _{p=0}^{g}=W_{C}^{g}\Gamma ^{C},\;\;\;\;\left( \frac{\partial \Sigma
^{g}}{\partial p^{\mu }}\right) _{p=0}=W_{\{\mu \}C}^{g}\Gamma
^{C}\;,\;\;\;\;\left( \frac{\partial \Sigma ^{g}}{\partial p^{\mu }\partial
p^{\nu }}\right) _{p=0}=\;W_{\left\{ \mu \nu \right\} C}^{g}\Gamma
^{C},\;\;\;\;  \label{def}
\end{equation}%
where we use the standard basis
\begin{equation}
\Gamma ^{C}:\;\Gamma ^{4}=I,\;\;\Gamma ^{\mu }=\gamma ^{\mu },\;\Gamma ^{\mu
\nu }=\sigma ^{\mu \nu },\;\;\Gamma ^{5}=\gamma ^{5}=i\gamma ^{0}\gamma
^{1}\gamma ^{2}\gamma ^{3},\;\;\Gamma ^{5,\mu }=\gamma ^{5}\gamma ^{\mu }.
\end{equation}%
This allows us to rewrite the self-energy as%
\begin{equation}
\Sigma ^{g}(p)=\left( W_{C}^{g}+p^{\mu }W_{\{\mu \}C}^{g}+\frac{1}{2}p^{\mu
}p^{\nu }W_{\left\{ \mu \nu \right\} C}^{g}\right) \Gamma ^{C}+O(p^{3}).
\label{SIGMAG}
\end{equation}%
In order to deal with the calculation of such coefficients it is convenient
to separate the modified photon propagator (\ref{LORGAUGE}) into its even and odd parts
\begin{equation}
{\Delta }_{\mu \nu }(k)={\Delta }_{\mu \nu }^{even}(k)+{\Delta }_{\mu \nu
}^{odd}(k),
\end{equation}%
and rewrite them in the more compact form
\begin{eqnarray}
\Delta _{\mu \nu }^{even}(k) &=&\eta _{\mu \nu }F_{1}+F_{2}\Big(|\mathbf{k|}%
^{2}\eta _{0\mu }\eta _{0\nu }-k_{l}k_{r}\delta _{\mu }^{l}\delta _{\nu
}^{r}\Big)={\Delta }_{\nu \mu }^{even}(k)={\Delta }_{\mu \nu
}^{even}(-k),\;\;\;\;\;\;  \label{EVENPROP} \\
\Delta _{\mu \nu }^{odd}(k) &=&iF_{3}k_{r}\epsilon ^{lmr}\eta _{\mu l}\eta
_{\nu m}=-{\Delta }_{\mu \nu }^{odd}(-k)=-{\Delta }_{\nu \mu }^{odd}(k),
\label{ODPROP}
\end{eqnarray}%
where%
\begin{equation}
F_{1} =-\frac{k^{2}}{\left( \left( k{^{2}}\right) ^{2}-4g^{2}|\mathbf{k|}%
^{2}k{_{0}^{4}}\right) },\qquad F_{2}=\frac{4g^{2}k{_{0}^{4}}/k^{2}}{\left(
\left( k{^{2}}\right) ^{2}-4g^{2}|\mathbf{k|}^{2}k{_{0}^{4}}\right) },\qquad
F_{3} = -\frac{2gk_{0}^{2}}{\left( \left( k{^{2}}\right) ^{2}-4g^{2}|%
\mathbf{k|}^{2}k{_{0}^{4}}\right) },  \label{F3}
\end{equation}%
are even functions of $\mathbf{k\;}$and $k{_{0}.}$

From the general expressions for the contributions in (\ref{SIGMAEXP}),
together with the symmetry properties of the propagator plus the symmetrical
integration over the three-momenta it is possible to determine that the
non-zero contributions to $\Sigma ^{g}(p)$ are
\begin{eqnarray}
\Sigma ^{g}(p) &=& A I+i\tilde{A}\;\gamma ^{i}\gamma ^{j}\gamma ^{k}\epsilon
^{ijk}+p^{0}B\gamma ^{0}-p^{i}\left( C\gamma ^{i}-i\tilde{C}\; \gamma
^{j}\gamma ^{k} \epsilon ^{ijk} \right)+\frac{1}{2} \left( p^{0}\right) ^{2}
\left(DI + i\tilde{D}\gamma ^{i}\gamma ^{j}\gamma ^{k}\epsilon ^{ijk} \right)
\nonumber \\
&&+ \frac{1}{2}\mathbf{p}^{2} \left(E I +i\tilde{E}\gamma ^{i}\gamma
^{j}\gamma ^{k}\epsilon ^{ijk} \right) +i\tilde{F}\;p^{0}p^{i}\left(\gamma
^{0} \gamma ^{j}\gamma ^{k} \epsilon ^{ijk} )+O(p^{3}\right).
\label{SIGMAEXP0}
\end{eqnarray}

We will be interested in analyzing only those terms that could give rise to
a finite and possibly unsuppressed LIV contribution when $g\rightarrow 0$.
In this limit we should recover QED, which is parity conserving so that we
know that the electron self energy must have the form.%
\begin{equation}
\Sigma ^{g=0}(p)=W_{0}I+W_{1}\left( p^{\mu }\gamma _{\mu }\right) +\frac{%
W_{3}}{2}p_{\mu }p^{\mu }I+O(p^{3}).  \label{SIGMAG0}
\end{equation}%
From this perspective all parity violating terms $\tilde{A},\tilde{C},\tilde{%
D}$ and $\tilde{E}$ in (\ref{SIGMAEXP0}) are subject to scrutiny and they
should be finally suppressed. On the other hand, the parity conserving
contributions can be rearranged in the following way
\begin{equation}
\Sigma _{+}^{g}(p)=AI+\left( B-C\right) p^{0}\gamma _{0}+C(p^{\mu }\gamma
_{\mu })+\frac{\left( D+E\right) }{2}\left( p^{0}\right) ^{2}I-\frac{E}{2}%
p_{\mu }p^{\mu }I+O(p^{3}),  \label{EVENPART}
\end{equation}%
so that according to our prescription we expect
\begin{equation}
\lim_{g\rightarrow 0}\left( B-C\right) =0,\qquad  \lim_{g\rightarrow
0}\left( D+E\right) =0,\qquad
\lim_{g\rightarrow 0}(A,\, C,\, -E\;)=(W_{0}, \, W_{1},\, W_{3}).
\end{equation}%

The general strategy to evaluate the required integrals is the following.
The structure of the denominators $\mathcal{D}$\ entering in them is of the
form
\begin{equation}
\mathcal{D}\mathcal{=}\left[ \left( k{^{2}}\right) ^{2}-4g^{2}|\mathbf{k|}%
^{2}k{_{0}^{4}+i\epsilon }\right] \left[ k^{2}-m^{2}+i\epsilon \right] ,
\end{equation}%
which can be rewritten
\begin{equation}
\mathcal{D}\mathcal{=}\left( 1-4g^{2}|\mathbf{k|}^{2}\right) \left[ k_{0}{%
^{2}{-}}\left( \omega _{-}^{2}\left( {\mathbf{k}}\right) {-i\epsilon }%
\right) \right] \left[ k_{0}{^{2}{-}}\left( \omega _{+}^{2}\left( {\mathbf{k}%
}\right) {-i\epsilon }\right) \right] \left[ k_{0}{^{2}-}\left( E^{2}\left(
\mathbf{k}\right) -i\epsilon \right) \right] .
\end{equation}%
Within the region of integration ($|\mathbf{k|}<1/\left( 2g\right) $), the
poles in the complex $k_{0}$ plane have the form
\begin{equation}
k_{01}=\mathcal{E}({|\mathbf{k|}})-i\epsilon ,\;\;k_{02}=-\mathcal{E}({|%
\mathbf{k|}})+i\epsilon,
\end{equation}%
with $\mathcal{E}({|\mathbf{k|}})>0$. Here $\mathcal{E}({|\mathbf{k}|})$
stands for any of the involved energies $\omega _{\pm }\left( \mathbf{k}%
\right)$ and $E\left( \mathbf{k}\right)=\sqrt{\mathbf{k}^2+m^2} $. In\ this
way it is always possible to perform a Wick rotation to the Euclidean
signature such that $k_{0}=ik_{4}$. Due to the remaining rotational
symmetry, together with the symmetrical integration over $\mathbf{k}$, one
is finally left with only two integration variables which are $k_{4}$ and $|%
\mathbf{k}|$ that can be conveniently rewritten in polar form.

\section{The LIV contributions}

In this section we present a detailed calculation of the corrections $%
W_{\{\mu \}C}^{g}$ to the electron self energy arising from the even sector
of the photon propagator ${\Delta}_{\mu \nu }$ corresponding to the $(B-C)$
term in Eq.\ (\ref{EVENPART}). The calculation of the remaining
contributions goes along similar lines and we only give the final results.

\subsection{General structure of the contributions}

As a first step it is convenient to split them into the following temporal
and spatial pieces%
\begin{equation}
\;\left( \frac{\partial \Sigma }{\partial p^{0}}\right)
_{p=0}=W_{\{0\}M}\Gamma ^{M}=-ie^{2}\int \frac{d^{4}k}{(2\pi )^{4}}\gamma
^{\mu }\left[ \frac{\gamma _{0}}{[k^{2}-m^{2}+i\epsilon ]}+\frac{%
2k^{0}(m-\gamma k)}{[k^{2}-m^{2}+i\epsilon ]^{2}}\right] \gamma ^{\nu }{%
\Delta }_{\mu \nu }(k)\,\mathcal{J}(k),  \label{firsttemp}
\end{equation}%
\begin{equation}
\left( \frac{\partial \Sigma }{\partial p^{i}}\right)
_{p=0}=W_{\{i\}M}\Gamma ^{M}\;=ie^{2}\int \frac{d^{4}k}{(2\pi )^{4}}\gamma
^{\mu }\left[ \frac{\gamma ^{i}}{[k^{2}-m^{2}+i\epsilon ]}-\frac{%
2k^{i}(m-\gamma k)}{[k^{2}-m^{2}+i\epsilon ]^{2}}\right] \gamma ^{\nu }{%
\Delta }_{\mu \nu }(k)\,\mathcal{J}(k).  \label{firstspacial}
\end{equation}%
and further separate each contribution according to the even and odd pieces
of the photon propagator ${\Delta }_{\mu \nu }(k)$. To simplify the notation
we have introduced
\begin{equation}
\mathcal{J}(k)=\mathcal{I}(k)\,\theta \left( \frac{1}{2g}-|\mathbf{k}%
{}|\right) .
\end{equation}

\subsubsection{The contributions from the even part of the propagator}

Taking the even part of the photon propagator in (\ref{firsttemp}), the
temporal component of the derivative is
\begin{equation}
\left( \frac{\partial \Sigma }{\partial p^{0}}\right)
_{p=0}^{even}=W_{\{0\}M}^{even}\Gamma ^{M}=-i\gamma ^{0}e^{2}\int \frac{%
d^{4}k}{(2\pi )^{4}}\left[ \frac{1}{(k^{2}-m^{2}+i\epsilon )}-\frac{%
2k_{0}^{2}}{(k^{2}-m^{2}+i\epsilon )^{2}}\right] \left( -2F_{1}\right) \,%
\mathcal{J}(k),  \label{EVEN0}
\end{equation}%
where we have used
\begin{equation}
\gamma ^{\mu }\gamma ^{0}\gamma ^{\nu }\Delta _{\mu \nu }^{even}=-2\gamma
^{0}F_{1}.
\end{equation}
The function $F_{1}$ was introduced in Eq. (\ref{F3})  and from (\ref%
{EVEN0}) we see that the only contribution is given by the component$%
\;W_{\{0\}0}^{even}$. Let us define the quantity
\begin{equation}
W_{\{0\}0}^{even}\equiv B=-2ie^{2}\int \frac{d^{4}k}{(2\pi )^{4}}\left[
\frac{1}{(k^{2}-m^{2}+i\epsilon )}-\frac{2k_{0}^{2}}{(k^{2}-m^{2}+i\epsilon
)^{2}}\right] \frac{k{^{2}}}{\left( \left( k{^{2}}\right) ^{2}-4g^{2}|%
\mathbf{k|}^{2}k{_{0}^{4}}\right) }\mathcal{J}(k).  \label{B}
\end{equation}%
The spatial contribution is (no sum over $i$)
\begin{equation}
\left( \frac{\partial \Sigma }{\partial p^{i}}\right)
_{p=0}^{even}=W_{\{i\}M}^{even}\Gamma ^{M}=-ie^{2} \int \frac{%
d^{4}k}{(2\pi )^{4}}\left[ \frac{1}{(k^{2}-m^{2}+i\epsilon )}+\frac{2\left(
k^{i}\right) ^{2}}{(k^{2}-m^{2}+i\epsilon )^{2}}\right] (\gamma ^{\mu }
\gamma ^{i}\gamma ^{\nu }\Delta _{\mu \nu }^{even})%
\mathcal{J}(k),
\end{equation}%
where we  use%
\begin{equation}
\gamma ^{\mu }\gamma ^{i}\gamma ^{\nu }\Delta _{\mu \nu }^{even}=2\gamma
^{i}\left( -F_{1}+(k_{i}^{2}-|\mathbf{k|}^{2})F_{2}\right).
\end{equation}
The rotational invariance of the three-momentum integration leads to
\begin{equation}
W_{\{i\}i}^{even}\equiv -C=2ie^{2}\int \frac{d^{4}k}{(2\pi )^{4}}\left[
\frac{1}{(k^{2}-m^{2}+i\epsilon )}+\frac{2|\mathbf{k|}^{2}/3}{%
(k^{2}-m^{2}+i\epsilon )^{2}}\right] \frac{\left( k{^{2}}-\frac{8g^{2}|%
\mathbf{k|}^{2}k{_{0}^{4}}}{3k^{2}}\right) }{\left( \left( k{^{2}}\right)
^{2}-4g^{2}|\mathbf{k|}^{2}k{_{0}^{4}}\right) }\mathcal{J}(k).  \label{C}
\end{equation}

\subsection{Calculation of the $(B-C)$\ contribution}

From Eqs. (\ref{B}) and (\ref{C}) we have
\begin{equation}
B-C=4ie^{2}\int \frac{d^{4}k}{\left( 2\pi \right) ^{4}}\frac{\left(
k_{0}^{2}+\frac{1}{3}\mathbf{k}^{2}\right) k{^{2}}-\frac{4g^{2}}{3k^{2}}|%
\mathbf{k|}^{2}k{_{0}^{4}}\left[ k^{2}-m^{2}+\frac{2}{3}\mathbf{k}^{2}\right]
}{\left( k^{2}-m^{2}+i\epsilon \right) ^{2}\left( \left( k{^{2}}\right)
^{2}-4g^{2}|\mathbf{k|}^{2}k{_{0}^{4}}\right) }\,\mathcal{J}(k).  \label{B-C}
\end{equation}%
In order to calculate the non covariant integrals of the above type,
together with those in the Appendix D, we give some details of the procedure
sketched at the end of the previous Section. Basically we implement the
following steps.

(i) First, we perform a Wick rotation to a Euclidean signature, such that
\begin{equation}
k_{0}=ik_{4},\;\;k^{2}=-(k_{4}^{2}+\mathbf{k}^{2})=-k_{E}^{2},\;\;%
\;d^{4}k_{E}=i\;4\pi |\mathbf{k|}^{2}\;dk_{4}\;d{|\mathbf{k|}}\;.\;
\end{equation}

(ii) Second, since we are maintaining rotational invariance we are left with
only two variables
\begin{equation}
-\infty <k_{4}<+\infty ,\;\;0<{|\mathbf{k|}}\;<\frac{1}{2g}.  \label{STRIP}
\end{equation}%
In this two-dimensional space we introduce the following polar coordinates
\begin{equation}
k_{4}=r\cos \alpha \;,\;\;\;\;\;{|\mathbf{k|=}}r\sin \alpha ,\;  \label{var}
\end{equation}%
where $k_{E}^{2}=r^{2}$. Next we have to integrate over the rectangular
strip defined by (\ref{STRIP}) and we choose first to integrate over $r$ and
subsequently over $\alpha $. In this way we have
\begin{equation}
\int d^{4}k=i \int d^{4}k_E=i\;4\pi \int_{0}^{\pi }d\alpha \sin ^{2}\alpha
\int_{0}^{1/(2g\sin \alpha )}r^{3}dr.
\end{equation}

Applying the above procedure to Eq. (\ref{B-C}) we have
\begin{equation}
B-C =-4e^{2}\int \frac{d^{4}k_{E}}{\left( 2\pi \right) ^{4}}\frac{\Big[%
k_{E}^{2}\left( k_{0E}^{2}-\frac{1}{3}\mathbf{k}^{2}\right) +\frac{4g^{2}}{%
3k_{E}^{2}}|\mathbf{k|}^{2}k{_{0E}^{4}}\left( -k_{E}^{2}-m^{2}+\frac{2}{3}|%
\mathbf{k|}^{2}\right) \Big]}{\left( k_{E}^{2}+m^{2}\right) ^{2}\left(
\left( k_{E}{^{2}}\right) ^{2}-4g^{2}|\mathbf{k|}^{2}k{_{0E}^{4}}\right) }
 \frac{M^{2}}{\left( M^{2}+k_{E}^{2}\right) }\theta \left( \frac{1}{%
2g}-|\mathbf{k}{}|\right) .
\end{equation}%
Introducing the polar coordinates (\ref{var}) yields
\begin{eqnarray}
B-C &=&-\frac{e^{2}}{\pi ^{3}}\int_{0}^{\pi }d\alpha \sin ^{2}\alpha
\int_{0}^{{1}/{(2g\sin \alpha )}}r^{3}dr\frac{\Big[(\cos ^{2}\alpha -\frac{1%
}{3}\sin ^{2}\alpha )+\frac{4}{3}g^{2}\sin ^{2}\alpha \cos ^{4}\alpha
(-r^{2}-m^{2}+\frac{2}{3}r^{2}\sin ^{2}\alpha )\Big]}{\left(
r^{2}+m^{2}\right) ^{2}\left( 1-4g^{2}r^{2}\sin ^{2}\alpha \cos ^{4}\alpha
\right) }  \nonumber \\
&&\times \frac{M^{2}}{\left( M^{2}+r^{2}\right) }.
\end{eqnarray}%
The required radial integrals are
\begin{equation}
K_{1}(\alpha )=\int_{0}^{1/(2g\sin \alpha )}\;\;\frac{r^{3}}{\left(
r^{2}+m^{2}\right) ^{2}\left( 1-r^{2}c^{2}(\alpha )\right) }\frac{M^{2}}{%
\left( M^{2}+r^{2}\right) }dr,  \label{K1}
\end{equation}%
\begin{equation}
K_{2}(\alpha )=\int_{0}^{1/(2g\sin \alpha )}\;\;\frac{r^{5}}{\left(
r^{2}+m^{2}\right) ^{2}\left( 1-r^{2}c^{2}(\alpha )\right) }\frac{M^{2}}{%
\left( M^{2}+r^{2}\right) }dr.  \label{K2}
\end{equation}%
which can be exactly calculated, yielding
\begin{equation}
K_{1}=\frac{M^{2}}{2}\left[ \frac{(c^{2}m^{4}+M^{2})}{\rho ^{2}(\Delta
^{2})^{2}}\ln \left( \frac{\Lambda _{m}^{2}}{m^{2}}\right) -\frac{M^{2}}{%
(\Delta ^{2})^{2}\eta }\ln \left( \frac{\Lambda _{M}^{2}}{M^{2}}\right) +%
\frac{c^{2}\ln (1-c^{2}b^{2})}{\rho ^{2}\eta }+\frac{b^{2}}{\rho \Lambda
_{m}^{2}\Delta ^{2}}\right] ,  \label{K1EXACT}
\end{equation}%
\begin{equation}
K_{2}=\frac{M^{2}}{2}\left[ -\frac{\left(
2m^{2}M^{2}+m^{4}(c^{2}M^{2}-1)\right) }{\rho ^{2}(\Delta ^{2})^{2}}\ln
\left( \frac{\Lambda _{m}^{2}}{m^{2}}\right) +\frac{M^{4}}{(\Delta
^{2})^{2}\eta }\ln \left( \frac{\Lambda _{M}^{2}}{M^{2}}\right) -\frac{\ln
(1-c^{2}b^{2})}{\rho ^{2}\eta }-\frac{b^{2}m^{2}}{\rho \Lambda
_{m}^{2}\Delta ^{2}}\right] .  \label{K2EXACT}
\end{equation}%
The notation is
\begin{eqnarray}
\rho  &=&1+m^{2}c^{2}(\alpha ),\;\;\;\eta =1+M^{2}c^{2}(\alpha ),\quad
\Delta ^{2}=m^{2}-M^{2},\quad c(\alpha )=2g\sin \alpha \cos ^{2}\alpha,
\nonumber \\
\Lambda _{m}^{2} &=&m^{2}+b^{2}(\alpha ),\quad \Lambda
_{M}^{2}=M^{2}+b^{2}(\alpha ),\quad b(\alpha )=\frac{1}{2g\sin \alpha }.
\end{eqnarray}%
In order to simplify the results by including only the dominant terms, we
will expand the above expressions in powers of $g^{2}$. This is justified
since the expressions (\ref{K1EXACT}) and (\ref{K2EXACT}) are free of poles.
Up to order $g^{2}$, the remaining integrals over $\alpha$ will be of the form%
\begin{equation}
\sin ^{p}\alpha \cos ^{q}\alpha ,\;\;\sin ^{p}\alpha \cos ^{q}\alpha \ln
(1-\cos ^{4}\alpha ),\;\;\sin ^{p}\alpha \cos ^{q}\alpha \ln (\sin \alpha
)\;,\;
\end{equation}%
with $p, q$ integers. These integrals contribute only with finite numerical
factors, which are not very relevant in order to establish the correct QED
limit of the LIV terms and only the final numerical results will be
presented. Nevertheless, we will isolate the exact $g^{2}$ independent
contribution and we will show that the angular integration produces a zero
contribution, thus eliminating any indication of fine-tuning. In all
the remaining contributions proportional to $g^{2}$ we will further expand
in powers of $m/M$ and retain only the dominant terms. In this way we
will need the approximate  expressions
\begin{eqnarray}
K_{1} &=&\Bigg(\frac{M^{4}}{(m^{2}-M^{2})^{2}}\ln \left( \frac{M}{m}\right) +\frac{%
M^{2}}{2(m^{2}-M^{2})}\Bigg) +2(gm)^{2}\sin ^{2}\alpha \cos ^{4}\alpha \Bigg(1+4\ln (2gm\sin \alpha )%
\Bigg) \nonumber \\
&&-2(gM)^{2}\sin ^{2}\alpha \Bigg(1+\cos ^{4}\alpha \left( 2\ln \left(
2gM\sin \alpha \right) -\cos ^{4}\alpha \ln \left( 1-\cos ^{4}\alpha \right)
\right) \Bigg),   \label{K1approx} \\
g^{2}K_{2}&=&-(gM)^{2}\Bigg(\ln (2Mg\sin \alpha )+\frac{1}{2}\ln (1-\cos
^{4}\alpha)\Bigg) +(gm)^{2}\Bigg(2\ln (2gm\sin \alpha )+\frac{1}{2}\Bigg).  \label{K2approx}
\end{eqnarray}%
It is important to observe that the exact $g^{2}$ independent term, contained in
the first bracket of Eq. (\ref{K1approx}) gives a
zero contribution in virtue of the angular integral factor
\begin{equation}
\int_{0}^{\pi }\;\sin ^{2}\alpha \;\left( \cos ^{2}\alpha -\frac{1}{3}\sin
^{2}\alpha \right) \;d\alpha =0.
\end{equation}%
Performing numerically the remaining angular integrations in the proposed
approximation we obtain
\begin{eqnarray}
B-C&=&\frac{e^{2}}{\pi ^{2}}\Bigg\{\frac{}{} (gM)^{2}\Big( -0.070+0.010\ln (gM)\Big)\nonumber \\
&&-(gm)^{2}\Big( 0.016+0.021\ln (gm)+0.031\ln \left( \frac{m}{M}\right) %
\Big) \Bigg\}. \label{FINALBMC}
\end{eqnarray}
The remaining contributions from the even sector are
\begin{equation}
A=\frac{e^{2}m}{\pi ^{2}}\Bigg( \frac{M^{2}}{2(m^{2}-M^{2})}\ln \left(
\frac{M}{m}\right) +(gM)^{2}\Big( 0.75+0.047\ln (gM)\Big)
-(gm)^{2}\Big(0.014 + 0.047\ln (gm)\Big) \Bigg), \label{AM}
\end{equation}
\begin{equation}
D+E=-\frac{e^{2}}{8\pi ^{2}}(g^{2}m)\left( \ln \left( \frac{m}{M}\right) -%
\frac{3}{4}\right).
\end{equation}
Finally, the odd contributions are
\begin{equation}
\tilde{A}=\frac{e^{2}}{6\pi ^{2}}g \Bigg( M^{2}\Big( 0.018+0.063\ln
(gM)\Big) -m^{2}\Big( 0.026+0.063\ln (gm)\Big) \Bigg), \qquad \tilde{C}
=\frac{e^{2}}{48\pi^{2}} (gm) \left(\ln\left(\frac{m}{M}\right)+\frac{1}{2} \right) ,
\end{equation}%
\begin{equation}
\tilde{D}=\tilde{F}=\frac{5ge^{2}}{48\pi ^{2}}\left( \ln \left( \frac{M}{m}\right) +%
\frac{4}{5}\right) , \qquad \tilde{E}=\frac{ge^{2}}{12\pi ^{2}}\left( \frac{13}{24}\ln \left( \frac{M}{m}%
\right) +\frac{17}{32}\right).  \label{E}
\end{equation}
The results obtained above, in the framework of our prescription to recover
QED, have precisely the expected property that reduce to zero when we turn
off the LIV correction parameterized by $g$, keeping $M$ fixed. Also, the
results are consistent with the fact that the unsuppressed contribution
which we still expect to diverge even after we set $g=0$ and subsequently $%
M\rightarrow\infty$, comes in the term $A$ written in Eq. (\ref{AM}). This
term corresponds precisely to the mass renormalization contribution in
standard QED.

\section{Microcausality violation}

In this section we provide an estimation of the microcausality violation
associated to our model. A comprehensive study of such violations is out of
the scope of the present work. Microcausality violation has been
previously studied in the fermionic sector of the Extended
Standard Model, for example \cite{LEHNERT}.

We work directly in the Coulomb gauge associated to our reference system
where $n^{\mu }=(1,\mathbf{0})$. We only consider points $x$ and $x^\prime$
which produce a space-like interval\ $(x-x^\prime)^2<0$. Unfortunately we
cannot perform a passive Lorentz transformation to reach the system where $%
x_{0}-x_{0}^\prime=0$, which might simplify the calculation. This is because
such transformation will change $n^\mu=(1, {\mathbf 0})$ into  $n^{\prime \mu }=$\ $\gamma (1,\mathbf{v})$
and then our system will turn out to be manifestly of the HOTD type, thus
requiring the application of the perturbative process of Ref.\ \cite%
{CHENGETAL}, which we have avoided in our particular reference frame.

Even in the standard QED case there is a drawback when working in the
Coulomb gauge, which is basically due to the apparent causality violation of
the theory arising from the instantaneous character of the scalar potential. When
dealing with the commutator $[A_{T}^{i}(x),\;A_{T}^{j}(x^{\prime })]$, which
is the naive starting point to test microcausality, this problem shows up
because this commutator is proportional to $\left( \delta ^{ij}-\frac{%
\partial ^{i}\partial ^{j}}{\nabla ^{2}}\right) D(x-x^{\prime })$ where $%
D(x-x^{\prime })$ is a function that has support only in the light cone.
Nevertheless, the operator $1/\nabla ^{2}$, which is just a shorthand for
the Green function $1/|\mathbf{r}-\mathbf{r}^{\prime }|$, acting upon $%
D(x-x^{\prime })$ produces non-zero results outside the light-cone, thus
yielding an apparent violation of microcausality. The canonical way of
dealing with this problem is to calculate the commutators of the gauge
invariant fields $\mathbf{E}$ and $\mathbf{B}$ for space-like separation. We
will follow the same route here and we will discuss only the commutator
\begin{equation}
\left[ \bar{\Pi}_{i}^{T}(x),\;\bar{\Pi}_{j}^{T}(x^{\prime })\right]
=-\partial _{0}^{2}\left[ D ^{ij}(x-x^{\prime })\right] \equiv \Omega
^{ij}(x-x^{\prime }),  \label{OMEGA}
\end{equation}%
which is the analogue of the electric fields commutator in standard QED, with
$\bar{\Pi}_{i}^{T}(x)$ been gauge invariant.
Here%
\begin{equation}
D ^{ij}(x-x^{\prime })=[\bar{A}_{T}^{i}(x),\;\bar{A}_{T}^{j}(x^{\prime
})].
\end{equation}
A direct calculation starting from Eq.(32) leads to
\begin{equation}
D ^{ij}(x-x^{\prime })=\int \frac{d^{3}\mathbf{k}}{(2\pi )^{3}}%
\,\sum_{\lambda }\frac{1}{2\omega _{\lambda }(\mathbf{k)}}\left( \varepsilon
^{i}(\lambda ,\mathbf{k})\varepsilon ^{j\ast }(\lambda ,\mathbf{k})e^{-i{k}%
_{\lambda }(x-x^{\prime })}-\varepsilon ^{i\ast }(\lambda ,\mathbf{k}%
)\varepsilon ^{j}(\lambda ,\mathbf{k})e^{i{k}_{\lambda }(x-x^{\prime
})}\right),
\end{equation}%
and we use $z^{\mu }=(x^{\mu }-x^{\mu \prime })$ in the sequel. Using the
relations (\ref{EIJ1}) and (\ref{EIJ2})  from the Appendix we arrive at
\begin{eqnarray}
D ^{ij}(z) &=&\sum_{\lambda }\int \frac{d^{3}\mathbf{k}}{(2\pi )^{3}}\,%
\frac{1}{2\omega _{\lambda }(\mathbf{k)}}\left[ \frac{1}{2}\left[ \delta
^{ij}-\frac{k^{i}k^{j}}{|\mathbf{k|}^{2}}\right] -\frac{i\lambda }{2}\left[
\epsilon ^{ijm}\frac{k^{m}}{|\mathbf{k|}}\right] \right] e^{-i\left( {\omega
}_{\lambda }z_{0}-\mathbf{k\cdot z}\right) }  \notag \\
&&-\sum_{\lambda }\int \frac{d^{3}\mathbf{k}}{(2\pi )^{3}}\,\frac{1}{2\omega
_{\lambda }(\mathbf{k)}}\left[\frac{1}{2}\left[ \delta ^{ij}-\frac{%
k^{i}k^{j}}{|\mathbf{k|}^{2}}\right] +\frac{i\lambda }{2}\left[ \epsilon
^{ijm}\frac{k^{m}}{|\mathbf{k|}}\right] \right] e^{i\left( {\omega }%
_{\lambda }z_{0}-\mathbf{k\cdot z}\right)},
\end{eqnarray}%
which can be rewritten as
\begin{eqnarray}
D ^{ij}(z) &=&\left[ \delta ^{ij}-\frac{\partial ^{i}\partial ^{j}}{|%
\mathbf{\nabla |}^{2}}\right] \frac{1}{2}\sum_{\lambda }\int \frac{d^{3}%
\mathbf{k}}{(2\pi )^{3}}\,\frac{1}{2|\omega _{\lambda }(\mathbf{k)|}}\left[
e^{-i\left( {\omega }_{\lambda }z_{0}-\mathbf{k\cdot z}\right) }-e^{i\left( {%
\omega }_{\lambda }z_{0}-\mathbf{k\cdot z}\right) }\right]  \notag \\
&&-\epsilon ^{ijm}\partial ^{m}\frac{1}{2}\sum_{\lambda }\int \frac{d^{3}%
\mathbf{k}}{(2\pi )^{3}}\,\frac{\lambda }{2|\mathbf{k|}|\omega _{\lambda }(%
\mathbf{k)|}}\left[ e^{-i\left( {\omega }_{\lambda }z_{0}-\mathbf{k\cdot z}%
\right) }-e^{i\left( {\omega }_{\lambda }z_{0}-\mathbf{k\cdot z}\right) }%
\right].  \label{DELTAF}
\end{eqnarray}%
Let us remark that this expression contains the correct limit when $g=0$. In
this case ${\omega }_{+}={\omega }_{+}=|\mathbf{k}|$, so that the
contributions of each term in the $\sum_{\lambda }$ are the same. After the
summation, the first line of (\ref{DELTAF})\ reproduces the definition of
the standard function $D(x-x')$, while the second line is proportional
to $\sum_{\lambda }\lambda =0$.

Starting from (\ref{OMEGA}) yields
\begin{eqnarray}
\Omega^{ij}(z) &=&-\left[ \delta ^{ij}\partial _{0}^{2}-\partial
^{i}\partial ^{j}\right] \frac{1}{2}\sum_{\lambda }\int \frac{d^{3}\mathbf{k}%
}{(2\pi )^{3}}\,\frac{\sqrt{1+2\lambda g|\mathbf{k|}}}{2|\mathbf{k|}}\left[
e^{-i\left( {\omega }_{\lambda }z_{0}-\mathbf{k\cdot z}\right) }-e^{i\left( {%
\omega }_{\lambda }z_{0}-\mathbf{k\cdot z}\right) }\right]  \notag \\
&&-\left( g\left[ \partial ^{i}\partial ^{j}\right] +\frac{1}{2}\epsilon
^{ijm}\partial ^{m}\right) \frac{1}{2}\sum_{\lambda }\int \frac{d^{3}\mathbf{%
k}}{(2\pi )^{3}}\,\left[ \frac{\lambda }{\sqrt{1+2\lambda g|\mathbf{k|}}}%
\right] \left[ e^{-i\left( {\omega }_{\lambda }z_{0}-\mathbf{k\cdot z}%
\right) }-e^{i\left( {\omega }_{\lambda }z_{0}-\mathbf{k\cdot z}\right) }%
\right] ,  \label{OMEGAF}
\end{eqnarray}%
where we have rearranged the above expression in such a way that the first
line of (\ref{OMEGAF}) recovers the standard QED result in the limit $%
g\rightarrow 0$, while the second line is equal to  zero. In this way the
microcausality violation is encoded in the functions%
\begin{eqnarray}
V_1(z) &=&\frac{1}{2}\sum_{\lambda }\int \frac{d^{3}\mathbf{k}}{(2\pi )^{3}}%
\,\frac{\sqrt{1+2\lambda g|\mathbf{k|}}}{2|\mathbf{k|}}\left[ e^{-i\left( {%
\omega }_{\lambda }z_{0}-\mathbf{k\cdot z}\right) }-e^{i\left( {\omega }%
_{\lambda }z_{0}-\mathbf{k\cdot z}\right) }\right],  \label{V1F} \\
V_2(z) &=&\frac{1}{2}\sum_{\lambda }\int \frac{d^{3}\mathbf{k}}{(2\pi )^{3}}%
\,\left[ \frac{\lambda }{\sqrt{1+2\lambda g|\mathbf{k|}}}\right] \left[
e^{-i\left( {\omega }_{\lambda }z_{0}-\mathbf{k\cdot z}\right) }-e^{i\left( {%
\omega }_{\lambda }z_{0}-\mathbf{k\cdot z}\right) }\right],  \label{V2F}
\end{eqnarray}%
which are now acted by local operators only.

Since we expect microcausality violations, we will estimate their impact
arising only from the function $V_1$. Notice that $V_1(z)=-V_1(-z)$ as can
be seen from the expression (\ref{V1F}) . After performing the angular
integrations we obtain%
\begin{equation}
V_1=\frac{1}{2(2\pi )^{2}}\frac{1}{ir}\sum_{\lambda }\int_{0}^{1/2g}\,dk%
\mathbf{\;}\frac{\sqrt{1+2\lambda gk}}{2}\left[ e^{-i\frac{k}{\sqrt{%
1+2\lambda gk}}z_{0}}-e^{i\frac{k}{\sqrt{1+2\lambda gk}}z_{0}}\right] \left[
e^{ikr}-e^{-ikr}\right],
\end{equation}%
where $k=|\mathbf{k|}$ and we have enforced the upper limit $1/2g$ in order
to have real frequencies $\omega _{\lambda }(k)$ according to Eq. (\ref{RELDISFIN}). The
spacelike character of the interval is written as
$
-r<z_{0}<r.
$
To proceed we introduce the phases%
\begin{equation}
\Phi _{1\lambda }(k)=k\left( r-\frac{1}{\sqrt{1+2\lambda gk}}z_{0}\right)
,\;\;\Phi _{2\lambda }(k)=k\left( r+\frac{1}{\sqrt{1+2\lambda gk}}%
z_{0}\right) ,  \label{PHASES1}
\end{equation}%
in terms of which we rewrite $V_1$ as%
\begin{equation}
V_1=\frac{1}{2(2\pi )^{2}}\frac{1}{ir}\frac{1}{2}\sum_{\lambda
}\int_{0}^{1/2g}\,dk\mathbf{\;}\sqrt{1+2\lambda gk}\;\Big[ e^{i\Phi
_{1\lambda }}+e^{-i\Phi _{1\lambda }}-e^{i\Phi _{2\lambda }}-e^{-i\Phi
_{2\lambda }}\Big].  \label{V1F1}
\end{equation}
In order to make an estimate of the region where microcausality violations
occur we concentrate in the calculation of the momentum integrals appearing
in Eq. (\ref{V1F1}). We apply the stationary phase method to the generic integral%
\begin{equation}
I_{\lambda }=\int_{0}^{1/2g}dk\;f_{\lambda }(k)e^{-i\Phi _{\lambda }\left(
k\right) },\;f_{\lambda }(k)=\sqrt{1+2\lambda gk},
\end{equation}%
where the relevant phases are given in Eq. (\ref{PHASES1}). The general
result for such integral is
\begin{equation}
I_{\lambda }=f_{\lambda }(\bar{k})e^{-i\Phi _{\lambda }\left( \bar{k}\right)
}\int_{0}^{1/2g}dk\;e^{-i\frac{1}{2}\left[ \frac{d^{2}\Phi }{dk^{2}}\right]
_{k=\bar{k}}(k-\bar{k})^{2}},
\end{equation}%
where $\bar{k}$ is the momenta that makes de phase stationary within the
interval $\left[ 0,1/2g\right] $.

We illustrate the calculation for the case of $\Phi_{1\lambda }$. The
remaining cases are completely similar and only the final results are
written. The exact expression for the momentum $\bar{k}$\ that extremizes $%
\Phi_{1\lambda }$ is given by the equation%
\begin{equation}
\frac{r}{z_{0}}=\frac{1+\lambda g\bar{k}}{\left( 1+2\lambda g\bar{k}\right)
^{\frac{3}{2}}}.  \label{GKEXTRE}
\end{equation}%
Observe that $\bar{k}$ appears always in the combination $g\bar{k}$ so
that the solution will be of the form
\begin{equation}
\bar{k}=\frac{1}{g}x\left(\frac{r}{z_{0}}\right),
\end{equation}%
where $x\left( \frac{r}{z_{0}}\right) $\ solves the corresponding equation
obtained from (\ref{GKEXTRE}). This is a complicated function of $\frac{r}{%
z_{0}}$ and to make some analytical progress the following approximation is
made. We found that in the range of $\frac{r}{z_{0}}=1+\epsilon$ with $%
\epsilon << 1$, the exact curve $x\left( \frac{r}{z_{0}}\right)$ is well
approximated by the straight line
\begin{equation}
\bar{k}_{1\lambda }=-\frac{\lambda }{2g}\left( \frac{r}{z_{0}}-1\right)
,\quad \frac{r}{z_{0}}>1,\;  \label{EXTRE1}
\end{equation}%
resulting from the expansion of the phase to order $k^{2}$ in Eq. (\ref%
{PHASES1}), which is
\begin{equation}
\Phi _{1\lambda }(k)=kr - \left(\allowbreak k-k^{2}\lambda g\right) z_{0}.
\end{equation}%
This means that we are considering a space-like region close to the
light-cone such that
\begin{equation}
(1-\epsilon)r<|z_{0}|<r.  \label{RANGEZ0}
\end{equation}%
A posteriori we will verify that our results in fact fall within the range
of the approximation. For this purpose it is convenient to rewrite the
condition (\ref{RANGEZ0}) by stating that the maximum allowed fractional
deviation $|\frac{\Delta z_{0}}{r}|$ has to satisfy
\begin{equation}
\left|\frac{\Delta z_{0}}{r}\right|< \epsilon.  \label{MAXDEVLIN}
\end{equation}
From now on it is convenient to separate the cases according to the sign of $%
z_{0}$. For $z_{0}>0$ the extremum (\ref{EXTRE1}) has to satisfy the
condition
\begin{equation}
0<\bar{k}_{1\lambda }=-\frac{\lambda }{2g}\left( \frac{r}{z_{0}}-1\right) <%
\frac{1}{2g}.
\end{equation}%
We observe that\ we have no solution for $\lambda =+1.\;$The choice $\lambda
=-1\;$requires
\begin{equation}
\frac{r}{2}<z_{0}.
\end{equation}%
In this way we have%
\begin{eqnarray}
\bar{k}_{1-} &=&\frac{1}{2g}\left( \frac{r}{z_{0}}-1\right) ,\qquad \Phi
_{1-}(\bar{k}_{1-})=\frac{\left( r-z_{0}\right) ^{2}}{4 gz_{0}},\qquad \frac{%
r}{2}<z_{0}<r,\;  \label{SOLTPOS1} \\
\sqrt{1+2\lambda g\bar{k}_{1-}} &=&\sqrt{2-\frac{r}{z_{0}}},\qquad \left[
\frac{d^{2}\Phi _{1-}(k)}{dk^{2}}\right] _{k=\bar{k}_{1-}}=-2gz_{0}.
\label{SOLTPOS2}
\end{eqnarray}%
The case $z_{0}<0$ produces
\begin{equation}
\Phi _{1\lambda }(k)=kr+\left( \allowbreak k-k^{2}\lambda g\right) |z_{0}|,
\end{equation}%
with%
\begin{equation}
\bar{k}_{1\lambda }^{\prime }=\frac{\lambda }{2g}\left( \frac{r}{|z_{0}|}%
+1\right) ,\qquad \frac{r}{|z_{0}|}>1.
\end{equation}%
The condition%
\begin{equation}
0<\bar{k}_{1\lambda }^{\prime }=\frac{\lambda }{2g}\left( \frac{r}{|z_{0}|}%
+1\right) <\frac{1}{2g}
\end{equation}%
cannot be satisfied neither for $\lambda =-1$, nor for $\lambda =+1$. The
former leads to negative $\bar{k}_{1\lambda }^{\prime }$, while the latter
requires $\frac{r}{|z_{0}|}<0$. In other words there is no solution for $%
z_{0}<0$.

The case of $\Phi _{2\lambda }(k)$ has solution only for $z_{0}<0$ and $%
\lambda =-1$. The results are%
\begin{eqnarray}
\bar{k}_{2-} &=&\frac{1}{2g}\left( \frac{r}{|z_{0}|}-1\right) ,\qquad \Phi
_{2-}(\bar{k}_{2-})=\frac{\left( r-|z_{0}|\right) ^{2}}{4 g|z_{0}|},\qquad%
\frac{r}{2}<|z_{0}|<r,\quad z_{0}<0\ ,\ \;  \label{SOLTNEG1} \\
\sqrt{1+2\lambda g\bar{k}_{2-}} &=&\sqrt{2-\frac{r}{|z_{0}|}},\qquad \left[
\frac{d^{2}\Phi _{2-}(k)}{dk^{2}}\right] _{k=\bar{k}_{2-}}=-2g|z_{0}|.
\label{SOLTNEG2}
\end{eqnarray}%
Substituting in (\ref{V1F1}) yields%
\begin{eqnarray}
V_1(z) &=&\frac{1}{4(2\pi )^{2}}\frac{1}{ir}\theta (z_{0})\sqrt{\frac{%
2z_{0}-r}{z_{0}}}\mathbf{\;}\left[ e^{i\frac{\left( r-z_{0}\right) ^{2}}{%
4gz_{0}}}\int_{0}^{1/2g}\,dk\;e^{-igz_{0}(k-\bar{k}_{1-})^{2}}\right]  \notag
\\
&&+ \frac{1}{4(2\pi )^{2}}\frac{1}{ir}\theta (z_{0})\sqrt{\frac{2z_{0}-r}{%
z_{0}}}\mathbf{\;}\left[ e^{-i\frac{\left( r-z_{0}\right) ^{2}}{4gz_{0}}%
}\int_{0}^{1/2g}\,dk\;e^{+igz_{0}(k-\bar{k}_{1-})^{2}}\right]  \notag \\
&&-\frac{1}{4(2\pi )^{2}}\frac{1}{ir}\theta (-z_{0})\sqrt{\frac{2z_{0}+r}{%
z_{0}}}\left[ e^{-i\frac{\left( r+z_{0}\right) ^{2}}{4gz_{0}}%
}\int_{0}^{1/2g}\,dke^{+igz_{0}(k-\bar{k}_{2-})^{2}}\mathbf{\;}\right]
\notag \\
&&-\frac{1}{4(2\pi )^{2}}\frac{1}{ir}\theta (-z_{0})\sqrt{\frac{2z_{0}+r}{%
z_{0}}}\left[ e^{+i\frac{\left( r+z_{0}\right) ^{2}}{4gz_{0}}%
}\int_{0}^{1/2g}\,dk\mathbf{\;}e^{-igz_{0}(k-\bar{k}_{2-})^{2}}\right] ,
\label{V1F1F}
\end{eqnarray}%
where we can verify that $V_1(z)=-V_1(-z)$. Though this will not be relevant
for our estimation of the microcausality violations, we can go one step further
and estimate the remaining integrals in the following way. Introducing the
change of variables\ $u=\sqrt{g}(k-\bar{k}_{1-})\;$we obtain%
\begin{equation}
I_{1\pm }=\int_{0}^{1/2g}\,dk\;e^{\pm igz_{0}(k-\bar{k}_{1-})^{2}}=\frac{1}{%
\sqrt{g}}\int_{-\sqrt{g}\bar{k}_{1-}}^{\sqrt{g}\left( \frac{1}{2g}-\bar{k}%
_{1-}\right) }\,du\;e^{\pm iz_{0}u^{2}}.
\end{equation}%
Substituting the value of $\bar{k}_{1-}$ results in
\begin{equation}
I_{1\pm }=\frac{1}{\sqrt{g}}\int_{-\frac{1}{2\sqrt{g}}\left( \frac{r}{z_{0}}%
-1\right) }^{\frac{1}{2\sqrt{g}}\left( 2-\frac{r}{z_{0}}\right)
}\,du\;e^{\pm iz_{0}u^{2}}\simeq \frac{1}{\sqrt{g}}\int_{-\infty }^{\infty
}\,du\;e^{\pm iz_{0}u^{2}}=\sqrt{\frac{\pi }{2gz_{0}}}(1\pm i).
\label{I1MASME}
\end{equation}

The expression for
\begin{equation}
I_{2\pm }=\int_{0}^{1/2g}\,dk\;e^{\pm igz_{0}(k-\bar{k}_{2-})^{2}}=%
\int_{0}^{1/2g}\,dk\;e^{\mp ig|z_{0}|(k-\bar{k}_{2-})^{2}}
\end{equation}%
can be obtained from (\ref{I1MASME})\ changing $z_{0}\;$by $|z_{0}|$, so
that we obtain%
\begin{equation}
I_{2\pm }=\sqrt{\frac{\pi }{2g|z_{0}|}}(1\mp i).
\end{equation}%
Then we have%
\begin{eqnarray}
V_1(z) &=&\frac{1}{4(2\pi )^{2}}\frac{1}{ir}\theta (z_{0})\sqrt{\frac{%
2z_{0}-r}{z_{0}}}\mathbf{\;}\sqrt{\frac{\pi }{2gz_{0}}}\left[ e^{i\frac{%
\left( r-z_{0}\right) ^{2}}{4gz_{0}}}(1-i)+e^{-i\frac{\left( r-z_{0}\right)
^{2}}{4gz_{0}}}(1+i)\right]  \notag \\
&&-\frac{1}{4(2\pi )^{2}}\frac{1}{ir}\theta (-z_{0})\sqrt{\frac{2z_{0}+r}{%
z_{0}}}\sqrt{\frac{\pi }{2g|z_{0}|}}\left[ e^{-i\frac{\left( r+z_{0}\right)
^{2}}{4gz_{0}}}(1-i)\mathbf{\;+}e^{+i\frac{\left( r+z_{0}\right) ^{2}}{%
4gz_{0}}}(1+i)\right].  \label{V1F1FF}
\end{eqnarray}

Next we analyze the regions where microcausality is violated and provide an
estimation of the amount of such violation. In our approximation such
violations occur when the functions $e^{\pm i\frac{\left( r-|z_{0}|\right)
^{2}}{4gz_{0}}}$ do not oscillate rapidly enough to make $V_1(z)$ equal zero
in the space-like region. Thus we take the condition for having
microcausality violations to be the region where the phases change slowly,
that is to say where
\begin{equation}
\frac{\left( r-|z_{0}|\right) ^{2}}{4g|z_{0}|}<1,  \label{MCVIOL}
\end{equation}%
in which case the oscillations are very much suppressed. Let us concentrate
now in the case $z_{0} > 0$ (the case $z_{0} < 0$ can be discussed in a
similar way). We first examine the curves that limit the region of interest
by considering the equality in Eq. (\ref{MCVIOL}). For a given $r$, the
solutions of such equation are
\begin{equation}
z_{0}{}_{+}=r+2g+2\sqrt{g^{2}+gr}, \qquad z_{0}{}_{-}=r+2g-2\sqrt{g^{2}+gr}.
\label{BOUNDARIES}
\end{equation}%
We observe that $z_{0}{}_{+}$ is always above the line $z_{0}=r$, while $%
z_{0}{}_{-}$ is always below. Also notice that both curves tend to the light
cone when $g\rightarrow 0$. The condition (\ref{MCVIOL}) is satisfied when
\begin{equation}
r-\left( 2\sqrt{g^{2}+gr}-2g\right) <z_{0}<r,  \label{REGION}
\end{equation}%
because this region includes the case $z_{0}\rightarrow r$ which clearly
satisfies\ (\ref{MCVIOL}). That is to say, (\ref{REGION}) determines the
space-like region where $V_1(z)$ is not zero, thus leading to microcausality
violations. For a given $r$, the range of $z_{0}$ within that region is
given by $\Delta z_{0}=r-z_{0-}$. Then we can quantify the maximum time
interval for which such violations occur by%
\begin{equation}
\frac{\Delta z_{0}}{|z_{0}|}\simeq \frac{\Delta z_{0}}{r}=\frac{1}{r}\left( 2%
\sqrt{g^{2}+gr}-2g\right) .  \label{FRACMCV}
\end{equation}%
The expression in the RHS of \ (\ref{FRACMCV}) is a monotonically decreasing
function of $r$ with the following end points
\begin{equation}
\left[ \frac{\Delta z_{0}}{r}\right] _{r\rightarrow 0}=1,\;\;\;\;\;\left[
\frac{\Delta z_{0}}{r}\right] _{r\rightarrow \infty }=0\;.
\end{equation}%
That is to say, for the whole region $r>r_{0}$ we can guarantee that
\begin{equation}
\frac{\Delta z_{0}}{r}<\left[ \frac{\Delta z_{0}}{r}\right] _{r=r_{0}}.
\label{UPPERLIM}
\end{equation}%
\begin{table}[tbp]
\begin{tabular}{|l|l|}
\hline
$\,\,\, r_0 [cm] \,\,\,$ & \, \, $\,\,\, \left| \Delta z_{0}\right| /r \, <
\, $ \\ \hline
\,\,\,1 & $\,\,\, 6.\,\allowbreak 3\times 10^{-22} $ \\
$\,\,\,10^{-10}$ & $\,\,\, 6.\,\allowbreak 3\times 10^{-17} $ \\
$\,\,\,10^{-22}$ & $\,\,\, 6.\,\allowbreak 3\times 10^{-11}$ \\
$\,\,\,10^{-27}$ & $\,\,\, 2.\,\allowbreak 0\times 10^{-8}$ \\
$\,\,\,10^{-33}$ & $\,\,\, 2.\,\allowbreak 0\times 10^{-5}$ \\ \hline
\end{tabular}%
\caption{Upper bound on fractional microcausality violation $\left| \Delta
z_{0}\right| /r $ for distances $r>r_0$}
\label{table1}
\end{table}
Recall that $g={\xi }/{\bar{M}}$, where $\xi$ is bounded by \ $10^{-10}$
when we choose\ $\bar{M}=M_{P}=10^{19}$ Gev ($L_P=10^{-33}$ cm) \cite%
{LIBERATI}. Thus, taking $g=10^{-43}$ cm and always considering the region $%
r>>g$, where we can trust the effective theory, we make some numerical
estimations of the relation (\ref{UPPERLIM}), which are given in Table \ref%
{table1}. The third and fourth values of Table \ref{table1} correspond to
distances given by $r_0= 10^{11} L_P $ and $r_0= 10^6 L_P$, which set a lower limit  beyond which
space becomes granular, according to the models considered in Refs. \cite%
{AMUFERMIONS} and \cite{KLINKHAMMER} respectively. The calculated
microcausality violations in Table I fall comfortably within the range
determined by (\ref{MAXDEVLIN}) required for the approximation to order $%
k^{2}$ in the phases (\ref{GKEXTRE} ) to be correct.

\section{Final Remarks}

In this work we have proposed a consistent quantization of the
electromagnetic sector of the Myers-Pospelov (MP) model, \cite{MP} coupled
to standard fermions, such that it can be realized as a perturbative
correction of standard QED. By this we mean that in the limit where the
Lorentz invariance violating (LIV) parameter $g=\xi/{\tilde M}$ goes to zero
one should recover the same quantum corrections arising in QED. Even though
this sector of the MP model is not of the higher order time derivative type,
up to a total derivative, some subtleties appear in the quantization of the
photon field. The correct perturbative prescription is achieved by
recognizing the effective character of the model via the introduction of a
coarse graining scale $M < < 1/g$, under which we assume that space retains
the usual attributes which allow the construction of a standard effective field
theory. Such cut-off scale is incorporated, in a smooth way,  by means of a Lorentz covariant
function of the Pauli-Villars type, which plays the role of a standard
regulator in the QED limit and makes sure that all LIV is codified in the
parameter $g$. The mathematical translation of this physical picture amounts
to the following prescription in order to properly recover QED: first take $%
g=0$, for constant $M$, and subsequently set $M \rightarrow \infty$. The
prescription has been tested in the calculation of LIV contributions arising
from the electron self energy, which indeed provide the expected results. In
this way the fine tuning problems found in Refs. \cite{GV,COLLINS} disappear
and one in fact recovers the correct zero limit for all the LIV corrections,
which are indeed shown to be very small perturbations in accordance with the
experimental and observational evidence.

Some comments regarding the plausibility of the scale $M$ in relation with
the very stringent constraints already found for LIV are now in order. The
combinations of parameters $g=\xi/{\bar M}, \,\, \eta_{1,2}/{\bar M} $, denoted
collectively by $\Xi / {\bar M} $, appearing in Eq. (\ref{MPLAG}) are
considered as remnants of a more fundamental quantum gravity (QG) theory,
which include effects that make space no longer describable in terms of a
continuum. Such parameters could arise, for example, in the process of
calculating expectation values of well defined QG operators in semiclassical
states that describe Minkowski space-time, which would be necessary to
derive the exact nature of the induced corrections to standard particle dynamics at low
energies. Let us emphasize that what is bounded by experiments or
observations is the ratio $\Xi /{\bar M} $, so that a neat separation of the
scale ${\bar M}$ and the correction coefficients $\Xi$, which could even be
zero if no corrections arise, is not possible until a semiclassical
calculation is correctly performed starting from a full quantum theory.
Initially, the naive expectation was that taking ${\bar M}=M_{Planck}$ will
be consistent with $\Xi $ values of order one, which is certainly not the
case. Nevertheless, we should not rule out rather unexpected values of $\Xi$
or ${\bar M}$ until the correct calculation is done.

Let us assume that we have identified the correct separation in ${\Xi_{QG}}/{%
M_{QG}}$ consistent with the experimental bounds and arising from a correct
semiclassical limit of the QG theory. Then we will interpret $M_{QG}$ as the
scale in which quantum effects are manifest and where space is characterized
by strong fluctuations forbidding its description as a continuum.
Nevertheless, another scale ${M}$ naturally should arise in this approach,
which is the one that separates the continuum description of space from a
foamy description related to quantum effects. That is to say, for probe
energies $E << {M}$ we are definitely within the standard continuum
description of space where effective field theory (EFT) methods should
apply. For probe energies $E >> {M} $ we enter the realm of quantum gravity
and there we assume that any EFT has to be replaced by an alternative
description. It is natural that a very large number of the basic quantum
cells of space characterized by the scale $\left( 1/M_{QG}\right)^{3}$ will
contribute to the much larger cells characterizing the onset of a continuum
description, so that we expect ${M} \ll M_{QG}$.

The maximum allowed momenta $|{\mathbf k}_{max}|\approx M_{QG}/\Xi_{QG}$ in the
theory will be mathematically dictated by the positivity of the normal modes
energies, Eq. (\ref{RELDISFIN}) in our case, and certainly constitutes an
extrapolation of the EFT that can be considered as the analogous of taking
the maximum momentum equal to infinity in the standard QED case. That is to
say, we need to introduce an additional suppression of the excitation modes
in our EFT which will be settled by the scale ${M}$, thus defining the
effective energy range of the model. This is required by the EFT description
of excitations in space which demands that the Compton wave length $1/|{\mathbf k}|$
of the allowed excitations be larger than the scale $1/{M}$ setting the
onset of the continuum. The implementation of this proposal is directly
related with our demand that the quantum model constructed from the MP
theory be such that it produces a continuous interpolation between those
physical results including $\Xi\neq 0$ corrections and those predicted by
standard QED ($\Xi=0$). In order to achieve this we have proposed the
prescription fully described in Section V.

Let us now discuss whether or not an estimate of the order of magnitude of
the scale $M$ in relation to $M_{QG}$ makes sense. In our specific case the
LIV contribution to the electron self energy produces an additional
dimension four contribution to the Lagrangian given by
\begin{equation}
\Delta L=\delta \frac{e^2}{\pi^2}\bar{\Psi}\gamma ^{0}i\partial _{0}\Psi,
\label{DELTAL}
\end{equation}
arising from the $(B-C)$ term in Eq. (\ref{EVENPART}). Our calculation leads
to a prediction dominated by
\begin{equation} |\delta| \sim 10^{-2}\times (gM)^2 |\ln(gM)|, \label{theorcorr}
\end{equation}
according to Eq. (\ref{FINALBMC}). On the other hand, starting from the
correction (\ref{DELTAL}) together with bounds from the anisotropy of the
inertial mass, the authors of Ref.\cite{GV} have established the
experimental bound
\begin{equation} |\delta|< 10^{-21}. \label{expcorr} \end{equation} In this way,
we expect that the scale $M$ is bounded in such a way that the theoretical correction
(\ref{theorcorr}) is much less than the experimental bound (\ref{expcorr}),
that is to say when
\begin{equation} g{M}=\xi \frac{M}{{\bar M}}=10^{-10}\frac{M}{{\bar M%
}} < < 0.65\times10^{-10}, \qquad \longrightarrow \qquad   M << 0.65 \, {\bar M}.\label%
{BOUND}
\end{equation}
The above shows that it is safe and consistent with present
observations to define a scale $M$ much below the quantum gravity scale ${%
\bar M}$.
Proposals for additional scales {$M$} significantly smaller than {${\bar M}%
=M_P $}, that can be understood as signaling the transition between the
standard space-time and that associated to the quantum gravity phase,
already exists in the literature. \cite{AMUFERMIONS,KLINKHAMMER}.

Next we comment upon the behavior of our result for the electron self-energy
under different momentum routings. For arbitrary internal momenta, the basic
expression (\ref{SELFENERGY}) can be rewritten as
\begin{equation}
\Sigma ^{g}(p)=-ie^{2}\int \frac{d^{4}k}{(2\pi )^{4}}\gamma ^{\mu
}S_F(k^\alpha+k^\alpha_2) \gamma ^{\nu }\bar{\Delta}_{\mu \nu
}(-k^\alpha+k^\alpha_1)\,\mathcal{I}(-k^\alpha+k^\alpha_1)\,\theta(1/(2g)-|{\mathbf{k_1-k}}%
|),  \label{SELFENERGY1}
\end{equation}%
where $k^\alpha_1-k^\alpha$ ($k^\alpha+k^\alpha_2$) is the internal photon
(electron) momentum respectively, with $k^\alpha_1$ and $k^\alpha_2$ being
arbitrary momenta satisfying the conservation $k^\alpha_2=p^\alpha-k^%
\alpha_1 $. Here $S_F(k^\alpha)$ denotes the standard fermion propagator.
Since the integral (\ref{SELFENERGY1}) is finite we are allowed to make the
change of variables $k^\alpha-k^\alpha_1 \longrightarrow k^\alpha $, which
reduces the integral to the form (\ref{SELFENERGY}) and shows its invariance
under momentum rerouting.

The stability of the model is guaranteed by restricting the observer Lorentz
covariance to concordant frames characterized by boosts factors up to $%
\gamma=1/\sqrt{2gM}$. Using the bound (\ref{BOUND}) the maximum allowed
boost factor is $\gamma_{max}=8.8\times 10^4$, which corresponds to a
maximum relative velocity such that $1-|\mathbf{v}_{max}|> 6.5\times10^{-11}$. This condition certainly includes concordant frames that move non-relativistically with respect to earth.

We have made a preliminary estimation of the microcausality violations in
the model by looking at the commutator of two gauge invariant momentum
operators (which are the extension of the electric field operators in
standard QED) for space-like separation $r>z_0$. The value of the
corresponding function has been calculated using the stationary phase
approximation and the condition for having microcausality violations
requires that the  exponentials oscillate very slowly. This
means that the associated phases should be of order one or less, which
defines a space-like region extremely close to the light cone, rapidly
approaching to it when the LIV parameter $g\rightarrow 0$. For a given value
of $r$ the width $|\Delta z_0|$ of such region is calculated. The fractional
value $(|\Delta z_0|/r)_{max}$ which sets the upper limit for the allowed
microcausality violation is subsequently estimated, leading to a typical
value of $|\Delta z_0|/r < 6.3\times 10^{-17}$ for distances $r$ larger than
the Compton wave length of the electron.

In this paper we have studied the construction of the quantum MP effective
model emphasizing the recovering of the correct QED limit in relation with
the absence of fine-tuning problems. A summary of our results has been presented
in Ref. \cite{MUV2}. Within the restrictions imposed we have established the basis
of a sound perturbative scheme to proceed with the calculation of additional
radiative processes. We defer for further work the analysis of the predictive
power of the model in relation to LIV corrections to physical observables.

\appendix

\section{}

\label{APPA}

The conventions used in this work for the Dirac algebra are those of Ref. %
\cite{Peskin} and we take $\hbar=c=1$. Also we have
\begin{eqnarray}
\eta _{\mu \nu } &=&diag(1,-1,-1,-1),\;\;\;\delta
^{ij}=+1,\;i=j,\;\;\;\delta ^{ij}=0,\;i\neq j,  \notag \\
\epsilon ^{0ijk} &=&\epsilon ^{ijk},\;\;\epsilon ^{0123}=\epsilon
^{123}=-\epsilon _{123}=+1,  \notag \\
\mathbf{A} &\mathbf{=}&\mathbf{(}A^{i}=-A_{i}\mathbf{),\;\;k=(}k^{i}=-k_{i}%
\mathbf{\ ),\;\;\;\nabla =}\left( \frac{\partial }{\partial x^{i}}=\partial
_{i}\right) ,  \notag \\
\partial _{r} &\Longleftrightarrow &-ik_{r},\;\;\partial
_{t}\Longleftrightarrow -ik_{0},\;\;\;\mathbf{A}\cdot \mathbf{B}%
=A^{i}B^{i},\;\;\;\left( \mathbf{A\times B}\right) ^{i}=\epsilon
^{ijk}A^{j}B^{k},\;\;\;\left( \mathbf{\nabla \times B}\right) ^{i}=\epsilon
^{ijk}\partial _{j}B^{k}.
\end{eqnarray}

In addition let us summarize some useful properties of the operator $\left(
M^{-1}\right) ^{ik}\;$introduced in Section \ref{secmodel},
\begin{equation}
\left( M^{-1}\right) ^{ji}\left( M^{-1}\right) ^{ir}=\frac{1}{\left(
1+4g^{2}\nabla ^{2}\right) ^{2}}\left[ \left( 1-4g^{2}\nabla ^{2}\right)
\delta ^{jr}-4g\epsilon ^{jpr}\partial _{p}+4g^{2}\left( 3+4g^{2}\nabla
^{2}\right) \partial _{j}\partial _{r}\right],  \label{I1}
\end{equation}%
\begin{equation}
\frac{1}{\left( 1+4g^{2}\mathbf{\nabla }^{2}\right) ^{2}}\left[ \left(
M^{-1}\right) ^{pi}\epsilon ^{ijk}\partial _{j}\left( M^{-1}\right) ^{kr}%
\right] =\left( 1-4g^{2}\nabla ^{2}\right) \epsilon ^{pjr}\partial
_{j}+4g\left( \delta ^{pr}\nabla ^{2}-\partial _{r}\partial _{p}\right),
\label{I3}
\end{equation}%
\begin{eqnarray}
\partial _{i}\partial _{j}\left( M^{-1}\right) ^{ij} &=&\frac{1}{\left(
1+4g^{2}\mathbf{\nabla }^{2}\right) }\left( \mathbf{\nabla }%
^{2}+4g^{2}\left( \mathbf{\nabla }^{2}\right) ^{2}\right) =\nabla ^{2},
\label{P1} \\
\partial _{j}\left( M^{-1}\right) ^{ij} &=&\frac{1}{\left( 1+4g^{2}\mathbf{%
\nabla }^{2}\right) }\left( 1+4g^{2}\mathbf{\nabla }^{2}\right) \partial
_{i}=\partial _{i},  \label{P2}
\end{eqnarray}%
\begin{equation}
\left( M^{-1}\right) ^{ij}\Pi _{j}^{T}=\frac{1}{\left( 1+4g^{2}\mathbf{%
\nabla }^{2}\right) }\left( \delta^{ij}+2g\epsilon^{irj}\partial _{r}\right)
\Pi _{j}^{T}.
\end{equation}

\section{}

\label{APPB}

Here we present the properties of the polarization vectors in the helicity basis ($\lambda=\pm 1$),
which are used in the expansion of the photon field in Eq.(\ref{AMUBAREXP}). They
satisfy the identities
\begin{equation}
\mathbf{\varepsilon }^{\ast }(\lambda ,\mathbf{\hat{k}}) =\mathbf{%
\varepsilon }(-\lambda ,\mathbf{\hat{k}}), \qquad
\mathbf{\hat{k}\cdot \;\varepsilon }(\lambda ,\mathbf{\hat{k}})=0, \qquad
\mathbf{\hat{k}\times \;\varepsilon }(\lambda ,\mathbf{\hat{k}})
=-i\lambda \mathbf{\varepsilon }(\lambda ,\mathbf{\hat{k}}),
\end{equation}
\begin{equation}
\mathbf{\varepsilon }^{\ast }(\lambda ,\mathbf{\hat{k}})\cdot \mathbf{%
\varepsilon }(\lambda ^{\prime },\mathbf{\hat{k}}) =\delta _{\lambda
\lambda ^{\prime }}, \qquad
\mathbf{\varepsilon }(\lambda ,-\mathbf{\hat{k}}) =\mathbf{\varepsilon }%
(-\lambda ,\;\mathbf{\hat{k}}), \qquad
\mathbf{\varepsilon }(\lambda ^{\prime },-\mathbf{\hat{k}})\cdot \left(
\mathbf{k}\times \mathbf{\varepsilon }(\lambda ,\mathbf{\hat{k}})\right)
=-i\lambda |\mathbf{k|}\delta _{\lambda \lambda ^{\prime }},
\end{equation}%
\begin{equation}
\varepsilon ^{r}(\lambda ,\mathbf{\hat{k}})\left[ \delta ^{rp}+i2g\epsilon
^{rmp}k_{m}\right] \varepsilon ^{p}(\lambda ^{\prime },-\mathbf{\hat{k}})=%
\left[ 1+2g\lambda |\mathbf{k|}\right] \delta _{\lambda \lambda ^{\prime }},
\label{ORT1}
\end{equation}%
\begin{equation}
\varepsilon ^{r\ast }(\lambda ,\mathbf{\hat{k}})\left[ \delta
^{rp}-i2g\epsilon^{rmp}k_{m}\right] \varepsilon ^{p}(\lambda ^{\prime },%
\mathbf{\hat{k}})=\left[ 1+2g\lambda |\mathbf{k|}\right] \delta _{\lambda
\lambda ^{\prime }}.  \label{ORT2}
\end{equation}%
The following combinations are useful in the construction of the
corresponding propagator
\begin{eqnarray}
\varepsilon ^{i}(\lambda ,\mathbf{\hat{k}})\varepsilon ^{j\ast }(\lambda ,%
\mathbf{\hat{k}}) &=&\frac{1}{2}\left[ \delta^{ij}-\frac{k^{i}k^{j}}{|%
\mathbf{k|}^{2}}\right] -\lambda \frac{i}{2}\left[ \epsilon ^{ijm}\frac{k^{m}%
}{|\mathbf{k|}}\right],  \label{EIJ1} \\
\varepsilon ^{i}(\lambda ,-\mathbf{\hat{k}})\varepsilon ^{j\ast }(\lambda ,-%
\mathbf{\hat{k}}) &=&\frac{1}{2}\left[ \delta^{ij}-\frac{k^{i}k^{j}}{|%
\mathbf{k|}^{2}}\right] +\lambda \frac{i}{2}\left[ \epsilon ^{ijm}\frac{k^{m}%
}{|\mathbf{k|}}\right],  \label{EIJ2}
\end{eqnarray}%
together with the sums%
\begin{equation}
\sum_{\pm \lambda }\varepsilon ^{i}(\lambda ,\mathbf{\hat{k}})\varepsilon
^{j\ast }(\lambda ,\mathbf{\hat{k}})=\sum_{\pm \lambda }\varepsilon
^{j}(\lambda ,-\mathbf{\hat{k}})\varepsilon ^{i\ast }(\lambda ,-\mathbf{\hat{%
k}})=\delta ^{ij}-\frac{k^{i}k^{j}}{|\mathbf{k|}^{2}}.
\end{equation}%
The final calculation of the propagator in Eq. (\ref{FINPROP0}) requires the
calculation of the following sums
\begin{eqnarray}
\sum_{\lambda }\frac{{1}}{{\left( k{_{0}^{2}-\omega _{\lambda
}^{2}+i\epsilon }\right) }} &=&\frac{2\left( k{^{2}}-4g^{2}|\mathbf{k|}^{2}k{%
_{0}^{2}}\right) }{\left( k{^{2}}\right) ^{2}-4g^{2}|\mathbf{k|}^{2}k{%
_{0}^{4}+i\epsilon }},  \label{S1} \\
\sum_{\lambda }\frac{{\omega _{\lambda }^{2}}}{{\left( k{_{0}^{2}-\omega
_{\lambda }^{2}}+i\epsilon \right) }} &=&\frac{{{2\mathbf{k}^{2}}}k^{2}}{%
\left( k{^{2}}\right) ^{2}-4g^{2}|\mathbf{k|}^{2}k{_{0}^{4}+i\epsilon }},\;
\label{S2} \\
\sum_{\lambda }\frac{{\lambda }}{{\left( k{_{0}^{2}-\omega _{\lambda }^{2}}%
+i\epsilon \right) }} &=&\;-\frac{4g|\mathbf{k|}^{3}}{\left( k{^{2}}\right)
^{2}-4g^{2}|\mathbf{k|}^{2}k{_{0}^{4}+i\epsilon }}.  \label{S3}
\end{eqnarray}

\section{}

\label{APPC}

The purpose of this Appendix is to compare the propagator obtained directly
from the equations of motion in the Lorentz gauge and further expressed in
the Coulomb gauge,  with the propagator (\ref{PROP1EXPL})\ obtained directly
in the Coulomb gauge after the canonical transformation (\ref{TCA}) is
made.

From the equations of motion (\ref{EQMOTLG}) in the Lorentz gauge \ we
identify the momentum space operator

\begin{equation}
O^{\nu \phi }(k)=-k^{2}\eta ^{\nu \phi }-2gik_{0}^{2}\epsilon ^{0\nu \sigma
\phi }k_{\sigma },  \label{OPLGMOM}
\end{equation}%
which propagator is
\begin{eqnarray}
\Delta _{\mu \nu }(k) &=&\frac{1}{(k^{2})^{2}-4g^{2}k_{0}^{4}\left| \mathbf{k%
}\right| ^{2})}\left[ -k^{2}\eta _{\mu \nu }+2igk_{0}^{2}\epsilon
^{pmq}k_{m}\eta _{p\mu }\eta _{q\nu }-\frac{4g^{2}k_{0}^{4}}{k^{2}}%
k_{p}k_{q}\delta _{\mu }^{p}\delta _{\nu }^{q}\right.  \notag \\
&&\left. +\frac{4g^{2}k_{0}^{4}{\left| \mathbf{k}\right| }^{2}}{k^{2}}\eta
_{0\mu }\eta _{0\nu }\right],  \label{PROPLGMOM}
\end{eqnarray}%
such that $O^{\nu \mu }\Delta _{\mu \rho }=\delta _{\rho }^{\nu }.\;\;$%
Separating the instantaneous Coulomb contribution
\begin{equation}
J^{i}(-k)\Delta _{ij}(k)J^{j}(k)=J^{\mu }(-k)\Delta _{\mu \nu }(k)J^{\nu
}(k)-J^{0}(-k)\frac{1}{|\mathbf{k|}^{2}}J^{0}(k)
\end{equation}
and using charge conservation \cite{Weinbergbook},
we find the corresponding propagator in the Coulomb gauge
\begin{equation}
\Delta _{ij}(k)=\frac{1}{|\mathbf{k|}^{2}((k^{2})^{2}-4g^{2}k_{0}^{4}\left|
\mathbf{k}\right| ^{2})}\left[ k^{2}\left( |\mathbf{k|}^{2}\delta
_{ij}-k_{i}k_{j}\right) +2gk_{0}^{2}|\mathbf{k|}^{2}i\epsilon ^{imj}k_{m}%
\right].  \label{PROPMARAT}
\end{equation}%
The Coulomb gauge propagator corresponding to the canonical transformation
(\ref{TCA}) and which was directly constructed from its vacuum expectation value
definition (\ref{phot-propg}) leading to the final result Eq. (\ref{PROP1EXPL}) is
\begin{equation}
\bar{\Delta}_{ij}=\frac{1}{|\mathbf{k|}^{2}\left( \left( k{^{2}}\right)
^{2}-4g^{2}k{_{0}}^{4}|\mathbf{k|}^{2}\right) }\left[ \left( k{^{2}}-4g^{2}|%
\mathbf{k|}^{2}k{_{0}^{2}}\right) \left( |\mathbf{k|}^{2}\delta
_{ij}-k_{i}k_{j}\right) +2g|\mathbf{k|}^{4}i\epsilon ^{imj}k_{m}\right].
\label{PROPLU}
\end{equation}%
Here we show the consistency between (\ref{PROPMARAT}) and (\ref{PROPLU}).

The starting point are the defining relations\
\begin{equation}
A=\Delta J_{T},\quad \bar{A}=\bar{\Delta}J_{T},
\end{equation}%
where both \ $A\;$and $\bar{A}\;$\ are in the Coulomb gauge. Whenever it is
not confusing we use the compact notation $A=\Delta J_{T}\Longleftrightarrow
A^{i}=\Delta _{ij}J_{T}^{j}$.\ \ Moreover, the photon fields  are related by the canonical
transformation $T$ such that
\begin{equation}
A=T\bar{A},\quad A^{\dagger }=\bar{A}^{\dagger }T^{\dagger }
\label{CTRANSF}
\end{equation}%
The invariant object is basically the electromagnetic energy
written in terms of the transverse current
\begin{equation}
E=\frac{1}{2}\int d^{3}x \, d^{3}y\;J_{T}^{i}(x)\Delta _{ij}(x-y)J_{T}^{j}(y),
\label{W}
\end{equation}%
where the sources are real. In momentum space this implies%
\begin{equation}
\left( J_{T}^{i}(k)\right) ^{\ast }=J_{T}^{i}(-k),
\end{equation}%
and Eq. (\ref{W}) translates into
\begin{equation}
E=\frac{1}{2}\int d^{3}k\;J_{T}^{i}(-k)\Delta _{ij}(k)J_{T}^{j}(k)=\frac{1}{2%
}\int d^{3}k\;J_{T}^{\dagger}\; \Delta \;J_{T}.
\end{equation}%
Reality of $E$ further demands
\begin{equation}
\Delta ^{\dagger}(k)=\Delta (k).
\end{equation}%
Expressing $E$ in terms of the fields yields
\begin{equation}
E=\frac{1}{2}\int d^{3}k\;J_{T}^{^{\dagger }}\Delta J_{T}=\frac{1}{2}\int
d^{3}k\;A^{\dagger }\;\left( \Delta ^{-1}\right) A,
\end{equation}%
in such a way that the equivalent description in terms of the barred
quantities \ requires
\begin{equation}
E=\frac{1}{2}\int d^{3}k\;\bar{A}^{\dagger }\;\left( \bar{\Delta}%
^{-1}\right) \;\bar{A}.
\end{equation}%
Inserting the transformation (\ref{CTRANSF}) we obtain the relation among
the two propagators
\begin{equation}
\Delta =T\bar{\Delta}T^{\dagger }.  \label{RELPROP}
\end{equation}%
The general structure of a Coulomb gauge propagator in our parity
violating theory can be written as
\begin{equation}
\Delta =\Delta _{1}K+\Delta _{2}S,\qquad \bar{\Delta}=\bar{\Delta}_{1}K+\bar{%
\Delta}_{2}S,   \label{COMPPROP}
\end{equation}%
where we have introduced the hermitian matrices
\begin{equation}
K=\left[ K_{ij}\right] =\left[ |\mathbf{k|}^{2}\delta _{ij}-k_{i}k_{j}\right]
,\;\;\;\;\;S=\left[ S_{ij}\right] =\left[ i\epsilon ^{imj}k_{m}\right] ,
\end{equation}%
with the following properties%
\begin{equation}
S^{3}=|\mathbf{k|}^{2}S,\;\;\;S^{2}=K,
\end{equation}%
in such a way that each propagator can be written in terms of the matrix $S$
only.

The canonical transformation $T$ has the form
\begin{equation}
T=\alpha I+\beta S=T^{\dagger },
\end{equation}%
in momentum space, with real numerical coefficients

\begin{equation}
\alpha =\frac{\sqrt{1+\sqrt{\left( 1+4a\right) }}}{\sqrt{2}\sqrt{\left(
1+4a\right) }},\qquad \beta =\;\frac{2g}{\sqrt{2}\sqrt{\left( 1+4a\right) }}%
\frac{1}{\sqrt{1+\sqrt{\left( 1+4a\right) }}},\qquad a=-g^{2}|\mathbf{k|}%
^{2}.
\end{equation}%
Substituting in the relation (\ref{RELPROP}) and after some algebra we
obtain the following conditions among the components of the respective
propagators
\begin{eqnarray}
\Delta _{1} &=&\frac{1}{\left( 1-4g^{2}|\mathbf{k|}^{2}\right) }\left[ \bar{%
\Delta}_{1}+2g\bar{\Delta}_{2}\right],  \label{RELCOMP1} \\
\Delta _{2} &=&\frac{1}{\left( 1-4g^{2}|\mathbf{k|}^{2}\right) }\left[ \bar{%
\Delta}_{2}+2g|\mathbf{k|}_{1}^{2}\bar{\Delta}\right].  \label{RELCOMP2}
\end{eqnarray}
From the expressions (\ref{PROPMARAT}) and (\ref{PROPLU})\ together with the
definition (\ref{COMPPROP})\ one can read%
\begin{eqnarray}
\Delta _{1} &=&\frac{k^{2}}{{|\mathbf{k|}^{2}}((k^{2})^{2}-4g^{2}k_{0}^{4}%
\left| \mathbf{k}\right| ^{2})},\qquad\Delta _{2}=\frac{2gk_{0}^{2}}{%
((k^{2})^{2}-4g^{2}k_{0}^{4}\left| \mathbf{k}\right| ^{2})}, \\
\bar{\Delta}_{1} &=&\frac{\left( k{^{2}}-4g^{2}|\mathbf{k|}^{2}k{_{0}^{2}}%
\right) }{|\mathbf{k|}^{2}\left( \left( k{^{2}}\right) ^{2}-4g^{2}k{_{0}}%
^{4}|\mathbf{k|}^{2}\right) },\qquad \bar{\Delta}_{2}=\frac{2g|\mathbf{k|}^{2}%
}{((k^{2})^{2}-4g^{2}k_{0}^{4}\left| \mathbf{k}\right| ^{2})},
\end{eqnarray}%
and verify that they satisfy the relations (\ref{RELCOMP1}) and (\ref%
{RELCOMP2}).

\section{}

\label{APPD}

In this Appendix we write the general form of the remaining LIV
contributions to the electron self energy which final results are presented
in Section VII, according to our general scheme of calculation
\begin{equation}
A=4ie^{2}m\int \frac{d^{4}k}{(2\pi )^{4}}\frac{(k^{2}-2g^{2}k_{0}^{4}|%
\mathbf{k|}^{2}/k^{2})}{[k^{2}-m^{2}]\left( \left( k{^{2}}\right)
^{2}-4g^{2}|\mathbf{k|}^{2}k{_{0}^{4}}\right) }\mathcal{J}(k),
\end{equation}%
\begin{equation}
\tilde{A}=\frac{2ie^{2}}{3}g\int \frac{d^{4}k}{(2\pi )^{4}}\frac{|\mathbf{k|}%
^{2}k_{0}^{2}}{[k^{2}-m^{2}]\left( \left( k{^{2}}\right) ^{2}-4g^{2}|\mathbf{%
k|}^{2}k{_{0}^{4}}\right) }\,\mathcal{J}(k),
\end{equation}%
\begin{equation}
\tilde{C}=\frac{4ie^{2}}{3}(gm)\int \frac{d^{4}k}{(2\pi )^{4}}\, \frac{|\mathbf{k|}%
^{2}k_{0}^{2}}{(k^{2}-m^{2})^{2}\left( \left( k{^{2}}\right) ^{2}-4g^{2}|%
\mathbf{k|}^{2}k{_{0}^{4}}\right) }\,\mathcal{J}(k),
\end{equation}%
\begin{equation}
D+E=32ie^{2}m\int \frac{d^{4}k}{\left( 2\pi \right) ^{4}}\frac{(k_{0}^{2}+|%
\mathbf{k|}^{2}/3)}{(k^{2}-m^{2})^{3}}\frac{(k^{2}-2g^{2}k_{0}^{4}|\mathbf{k|%
}^{2}/k^{2})}{\left( \left( k{^{2}}\right) ^{2}-4g^{2}|\mathbf{k|}^{2}k{%
_{0}^{4}}\right) }\mathcal{J}(k),
\end{equation}%
\begin{equation}
\tilde{D}=\tilde{F}=-\frac{4ie^{2}}{3}g\int \frac{d^{4}k}{(2\pi )^{4}}\left[
\frac{1}{(k^{2}-m^{2})^{2}}-\frac{4k_{0}^{2}}{(k^{2}-m^{2})^{3}}\right]
\frac{|\mathbf{k|}^{2}k_{0}^{2}}{\left( \left( k{^{2}}\right) ^{2}-4g^{2}|%
\mathbf{k|}^{2}k{_{0}^{4}}\right) }\,\mathcal{J}(k),
\end{equation}%
\begin{equation}
\tilde{E}=\frac{4ige^{2}}{3}\int \frac{d^{4}k}{(2\pi )^{4}}\, \left[ \frac{3}{%
(k^{2}-m^{2})^{2}}+\frac{\frac{4}{3}|\mathbf{k|}^{2}}{(k^{2}-m^{2})^{3}}%
\right] \frac{|\mathbf{k|}^{2}k_{0}^{2}}{\left( \left( k{^{2}}\right)
^{2}-4g^{2}|\mathbf{k|}^{2}k{_{0}^{4}}\right) }\mathcal{J}(k).
\end{equation}

\begin{acknowledgments}
C. M. R. acknowledges support from DGAPA-UNAM through a posdoctoral fellowship. L.F.U is partially
supported by projects CONACYT \# 55310 and DGAPA-UNAM-IN109107. J. D. V acknowledges
support from the projects CONACYT \# 47211-F and DGAPA-UNAM-IN109107.
\end{acknowledgments}

\end{document}